\documentclass[aps,twocolumn,superscriptaddress,floatfix,prx,longbibliography]{revtex4-1}
\usepackage[utf8]{inputenc}
\usepackage{lipsum}
\usepackage{graphicx}
\usepackage{amsmath,amssymb}
\usepackage{braket}
\usepackage{mathtools}
\usepackage{hyperref}
\usepackage{multirow}
\usepackage{color}
\usepackage[normalem]{ulem}
\usepackage{amsfonts}
\usepackage{bbm}
\usepackage[ruled,vlined]{algorithm2e}
\usepackage{braket}

\renewcommand{\vec}[1]{\boldsymbol{#1}} 

\makeatletter

\usepackage{xcolor}

\usepackage[normalem]{ulem} 

\usepackage[caption=false]{subfig}
\newcommand{\phantomsubfloat}[1]{
    {
        \captionsetup[subfigure]{labelformat=empty}
        \subfloat[][]{#1}
    }%
}

\AtBeginDocument{\let\LS@rot\@undefined}
\makeatother

\begin{document}
\newcommand{\linkstate}[3]{(\ensuremath{#1#2;#3})}
\newcommand{\state}[1]{\ensuremath{| {#1} \rangle}}
\newcommand{\mindex}{\ensuremath{\mathbf{m}}}
\newcommand{\mindexq}{\ensuremath{{\{\vec{m_q}\}}}}
\newcommand{\avg}[1]{\ensuremath{\left[ #1 \right] }}

\title{Generalized hydrodynamics: a perspective}

\author{Benjamin Doyon}
\affiliation{Department of Mathematics, King’s College London, Strand, London WC2R 2LS, UK}

\author{Sarang Gopalakrishnan}
\affiliation{Department of Electrical Engineering, Princeton University, Princeton, NJ 08544, USA}

\author{Frederik M\o ller}

\author{J\"org Schmiedmayer}
\affiliation{Vienna Center for Quantum Science and Technology, Atominstitut, TU Wien, Stadionallee 2, A-1020 Vienna, Austria}

\author{Romain Vasseur}
\affiliation{Department of Physics, University of Massachusetts, Amherst, MA 01003, USA}

\begin{abstract}

Conventional hydrodynamics describes systems with few long-lived excitations. In one dimension, however, many experimentally relevant systems feature a large number of long-lived excitations even at high temperature, because they are proximate to integrable limits. Such models cannot be treated using conventional hydrodynamics. The framework of generalized hydrodynamics (GHD) was recently developed to treat the dynamics of one-dimensional models: it combines ideas from integrability, hydrodynamics, and kinetic theory to come up with a quantitative theory of transport. GHD has successfully settled several longstanding questions about one-dimensional transport; it has also been leveraged to study dynamical questions beyond the transport of conserved quantities, and to systems that are not integrable. In this article we introduce the main ideas and predictions of GHD, survey some of the most recent theoretical extensions and experimental tests of the GHD framework, and discuss some open questions in transport that the GHD perspective might elucidate.

\end{abstract}
\maketitle

\setcounter{tocdepth}{1}

\tableofcontents

\section{Introduction}

Hydrodynamics is one of the pillars of modern many-body physics~\cite{2015arXiv151103646C,
2018arXiv180509331G,2017arXiv170208894L,PhysRevB.97.035127,kvh, rpv,PhysRevLett.122.091602,PhysRevX.9.041017,PhysRevResearch.2.033124,PhysRevB.105.075131}. At its core, hydrodynamics is an effective field theory description for the transport of extensive conserved quantities 
or other slow modes
in strongly interacting systems.
Hydrodynamics is built around the assumption that every degree of freedom that is not required to be slow by symmetry undergoes rapid local relaxation. 
%
This assumption implies that the current of any conserved quantity (and indeed any local observable) can be expressed as a gradient expansion in the conserved densities, such as the energy and particle density. 
It follows that the full dynamics is well described by an emergent, closed set of partial differential equations in terms of the conserved densities---typically a smaller set of degrees of freedom.
%
Hydrodynamics applies in a variety of contexts, from traditional classical fluid dynamics to ultracold quantum gases~\cite{sommer2011universal,cao2011universal}, quark-gluon plasmas~\cite{romatschke_romatschke_2019}, black hole physics~\cite{Rangamani_2009}, or electron fluids in pristine solid state systems with strong interactions and long mean free times such as graphene and PdCoO$_2$~\cite{Bandurin1055,Crossno1058,Moll1061,Sulpizio:2019aa}. Hydrodynamics is particularly rich for low-dimensional systems, featuring transport anomalies such as long-time tails~\cite{PhysRevLett.18.988,Ernst:1984aa,PhysRevB.73.035113,PhysRevA.89.053608, spohn1991large} and superdiffusion~\cite{PhysRevLett.78.1896,spohn_nlfhd,Spohn2014,delacretaz2020breakdown}.

Conventional hydrodynamics becomes intractable when there are too many slow degrees of freedom. This situation arises in one-dimensional integrable systems~\cite{Takahashi,mussardo2010statistical,baxter2016exactly,1742-5468-2016-6-064001,Bastianello_2022}, but is more general: emergent symmetries are characteristic of Fermi liquids at low temperature~\cite{PhysRevX.11.021005}, and also occur in the contexts of many-body localization and prethermalization~\cite{gopalakrishnan2020dynamics}. 
%
%
%
Here we will focus on integrable systems, as these remain the best-understood case. 
Although integrability might seem fine-tuned, many paradigmatic models of quantum many-body physics, such as the Hubbard, Heisenberg, and Lieb-Liniger models, are integrable. 
Integrability was initially seen as a computational tool, allowing for exact calculations of equilibrium properties~\cite{Takahashi,mussardo2010statistical}, but since the advent of cold-atom experiments~\cite{kinoshita} it has become clear that integrability has striking dynamical consequences~\cite{eisert2010colloquium,pssv}.
%
Yet, frustratingly, traditional exact solutions have been of limited impact to tackle far-from-equilibrium dynamics or even to compute linear-response conductivities. Computing dynamical quantities requires knowing the matrix elements of physical observables between arbitrary eigenstates, and summing over all the eigenstates of the system. Neither task seemed feasible prior to the recent developments that we survey here. 

In 2016, two groups independently came up with what is now known as generalized hydrodynamics (GHD)~\cite{Doyon, Fagotti}. 
GHD extends the general principles of hydrodynamics to the integrable setting, where there are extensively many conserved quantities. 
%
%
The main technical breakthrough~\cite{Doyon, Fagotti} was to find a basis in which the extensively many resulting hydrodynamic constitutive equations could be written in a way that remains not only mathematically tractable but intuitive:
%
the GHD equations are in essence collisionless Boltzmann equations for the stable quasiparticle excitations of integrable systems. 

Since then, GHD has been understood to form a universal framework for the emergent large-scale dynamics of a wide array of many-body integrable systems~\cite{Bastianello_2022}, from classical to quantum~\cite{Doyon, Fagotti,Alba_2021}, and from particles~\cite{doyon2017dynamics,PhysRevLett.120.164101} and solitons~\cite{Bonnemain_2022,El_2021} to spins~\cite{Gopalakrishnan_2023} and fields~\cite{Doyon}. GHD has led to a quantitative understanding of far-from-equilibrium transport setups, shed light on entanglement dynamics~\cite{alba2017entanglement,Bertini_2018,10.21468/SciPostPhys.7.1.005} and correlation spreading~\cite{10.21468/SciPostPhys.5.5.054,10.21468/SciPostPhysCore.3.2.016}, and provided analytical expressions for linear response quantities such as Drude weights~\cite{PhysRevLett.119.020602,BBH, GHDII, PhysRevB.96.081118, PhysRevB.97.081111,nagy2023thermodynamics}  and diffusion constants~\cite{dbd1, ghkv, dbd2,GV19,10.21468/SciPostPhys.9.5.075,doyon2019diffusion}. GHD also uncovered new regimes of anomalous transport in strongly interacting spin chains~\cite{PhysRevLett.106.220601,lzp,idmp,GV19,PhysRevLett.123.186601,gvw, dupont_moore,vir2019, 2019arXiv190905263A,PhysRevLett.122.210602,PhysRevB.102.115121,2020arXiv200908425I,dmki,PhysRevLett.125.070601}, and provides a starting point to study near-integrable systems. In turn, these theoretical developments motivated new experiments~\cite{PhysRevLett.122.090601,Jepsen:2020aa,2020arXiv200906651M,2020arXiv200913535S,PhysRevLett.126.090602,doi:10.1126/science.abf0147,wei2022quantum,Bouchoule_2022,hydrodynamization, PhysRevX.12.041032}, and GHD quantitatively predicts dynamics observed in experimental systems to high accuracy.

This perspective article aims to summarize and contextualize six years of rapid progress in understanding the dynamics of integrable systems, based on GHD. Our aim is not to present a detailed exposition of GHD: instead, we will provide a bird's-eye view of GHD, and focus on how GHD relates to open questions in nonequilibrium dynamics in diverse physical contexts.  For more details, we refer the reader to recent pedagogical introductions to generalized hydrodynamics~\cite{doyon2019lecture,Bastianello_2022,Bouchoule_2022,ESSLER2022127572}; and more specifically: for diffusive corrections to GHD, see~\cite{De_Nardis_2022}; for an introduction to integrability-breaking perturbations within GHD see~\cite{Bastianello_2021}. The review article~\cite{bertini2020finite} has an extensive discussion of numerical and exact results on transport in one-dimension systems. 

\section{Generalized hydrodynamics: general framework}

\subsection{Context}

The resurgence of interest in integrable dynamics is due in large part to a 2006 experiment on the far-from-equilibrium dynamics of cold atomic gases confined to move in a one-dimensional harmonic trap~\cite{kinoshita}. The gas was taken out of equilibrium by laser pulses that sent half of it flying to the left and the other half to the right. After climbing the confining potential well, the two halves of the gas returned and collided repeatedly, as in a Newton's cradle toy, with strong (and tunable) contact interactions. Remarkably, these collisions did not cause the momentum distribution of the gas to relax, in conflict with the naive expectation from statistical mechanics: instead, the particles seemed to ``go through each other'' and retain their initial momentum distribution. A natural way to reconcile these observations was that the experimental system was approximately described by the integrable Lieb-Liniger model~\cite{PhysRev.130.1605,PhysRev.130.1616}. Could it be, the authors asked, that this ``quantum Newton's cradle" effect was simply a consequence of integrability? While suggestive, this idea raised many questions, such as why integrability would conserve the distribution of individual particles' momenta, and why it would survive in the presence of a trap.

This experiment sparked a wealth of theoretical studies in the non-equilibrium dynamics of integrable systems. First addressed was the simplified, more basic question (see, e.g., Refs.~\cite{1742-5468-2016-6-064001}) of how integrable systems relax.
%
It was found that after quantum quenches---where from the ground state a parameter of the model is suddenly modified---the state of an integrable system relaxes, not to a Gibbs states, but to a generalized Gibbs ensemble (GGE)~\cite{PhysRevLett.98.050405,Rigol:2008kq,1742-5468-2016-6-064007}. These still take the Gibbs form $e^{-\sum_i \beta_i Q_i}$, but this involves in general the full, infinite family $\{Q_i\}$ of conserved quantities admitted by the system. The problem of organising and determining the set of $Q_i$'s led to surprising developments in many-body integrability~\cite{1742-5468-2016-6-064001}. For certain quenches it was possible, by ground-breaking technical Bethe ansatz calculations, to evaluate exactly the GGE appearing at long times in systems of infinite volumes~\cite{caux_essler,caux_konik,Caux_2016}. This involved new formulae for exact overlaps between such states and Bethe ansatz states. Overall, these studies made subtleties of relaxation more evident: it is quantum entanglement that gives rise, on finite regions, to relaxation to GGEs; and the order of limits is crucial.

However, the linear response theory and the novel developments in quantum quenches, although technically impressive, did not help in solving the original problem of the quantum Newton cradle. 
In such experiments, in general, the system is not only far from equilibrium, but also inhomogeneous. One loses both time and space stationarity. Re-constructing ground states for inhomogeneous potentials in terms of Bethe ansatz states is numerically possible but extremely expensive, and thermal states are even less accessible; special Bethe ansatz formulae for overlaps appear not to be available in general (although certain examples have been worked out, see Refs~\cite{De_Nardis_2018,CortesCubero2019,10.21468/SciPostPhys.8.1.004, G_hmann_2017, Kitanine_2011} and Ref.~\cite{Cubero_2021} and references therein for a recent review of form factors in the context of GHD). Reconstructing the evolution Hamiltonian with an external inhomogeneous potential in terms of Bethe ansatz states is likewise extremely difficult. The known, legacy techniques of quantum integrability appeared to be powerless to describe, let along explain, the far-from-equilibrium dynamics of atomic clouds in near-integrable systems.

\subsection{Euler scale generalized hydrodynamics}

\begin{figure}[t!]
	\centering 
	\includegraphics[width=.9\linewidth]{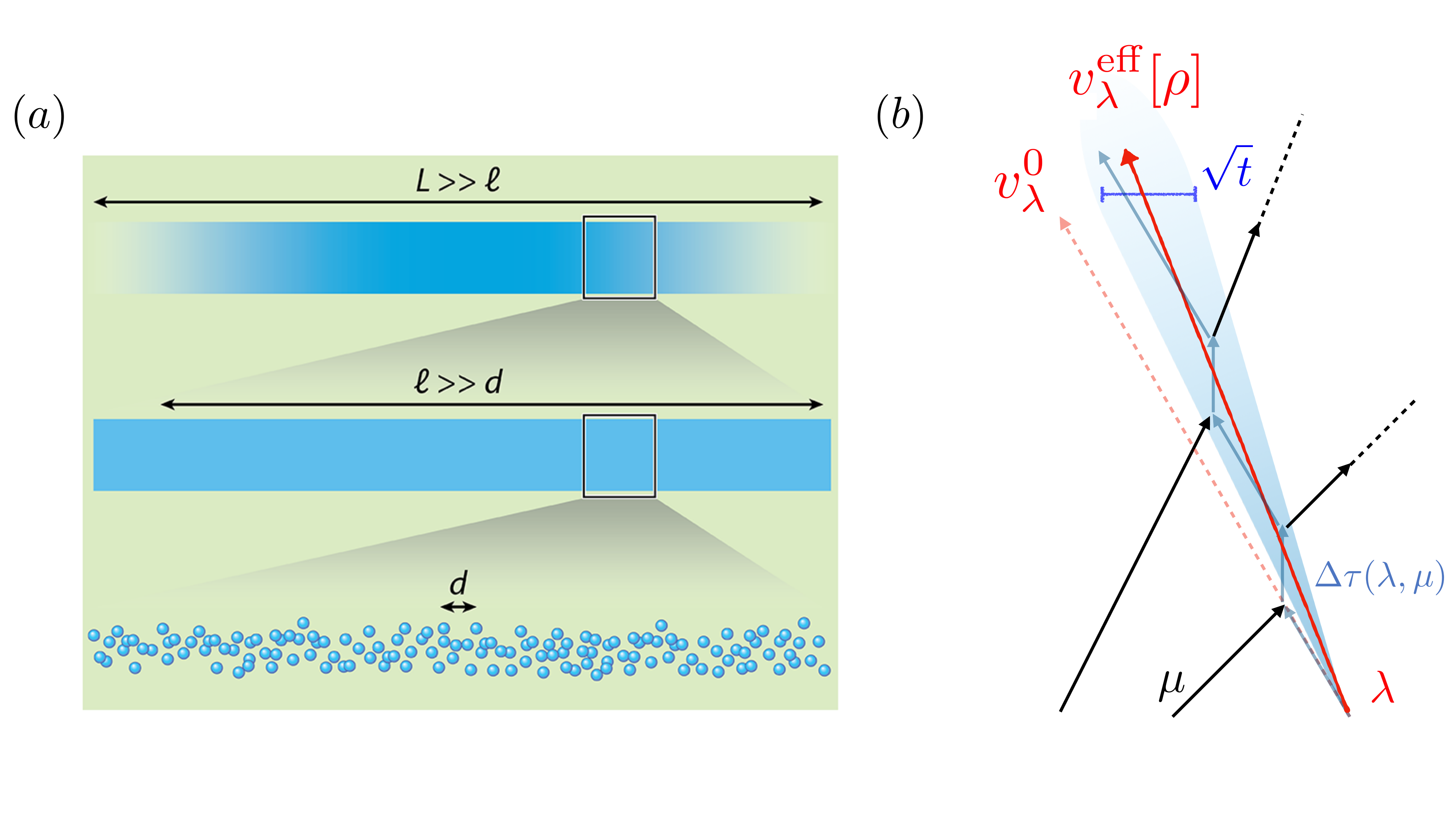}
	\caption{{\bf Hydrodynamics and quasiparticles.}  (a) Generalized hydrodynamics is an effective description of many-body integrable quantum systems on mesoscopic scales $\ell$ much larger than the lattice spacing or inter-particle distance $d$, but much smaller than the total system size $L$, describing the evolution from local to global equilibrium. Reprinted figure with permission from Ref.~\onlinecite{JeromePhysics}, Copyright 2016 by the American Physical Society.  (b) Soliton gas picture: a quasiparticle with rapidity $\lambda$ and bare velocity $v_\lambda^0$ moves on average with a renormalized velocity $v^{\rm eff}_\lambda$ due to time delays caused by collisions with other quasiparticles. Thermal fluctuations in the initial state leads to diffusive corrections to this leading ballistic motion.   }
	\label{fig:hydro}
\end{figure}

A key feature of integrable systems is the existence of an extensive number of extensive conserved quantities $Q_n = \int dx\, q_n(x)$, with associated continuity equations for their (local or quasi-local) densities and currents,
\begin{equation}\label{conservation}
\partial_t q_n + \partial_x j_n=0.
\end{equation}
The main idea of GHD is to follow the general logic of hydrodynamics, but taking into account all these conserved quantities. Imagine initializing a system in a non-equilibrium  state, and partitioning the system into local mesoscopic ``hydrodynamic cells.'' After a (supposedly) short local thermalization time, the local state in each cell relaxes. This means that it approaches a statistical ensemble determined by the relevant conserved quantities: in non-integrable systems, a Gibbs ensemble; in integrable systems, a generalized Gibbs ensemble. Thus the system's overall state at any time after local relaxation is simply fixed by the cell-dependent values of all conserved densities at that time. Since all other information about the initial state has been lost, the subsequent evolution just involves the dynamics, and eventual equilibration, of these few remaining slow variables across the system (Fig.~\ref{fig:hydro}a). Looking for an asymptotic expansion in the long-wavelength variations of these variables, one can write down the most general equations for these relaxation processes within a gradient expansion. 

In particular, the currents $j_n$ can be decomposed into ``fast'' components, which quickly decay, and ``slow'' components associated with their projections onto conserved densities and their gradients. At the longest wavelengths---the Euler scale---with $q_n$'s the values of the conserved densities in a given fluid cell, local relaxation gives the currents $j_n = \j_n[q]$ that would be obtained in the corresponding (generalized) Gibbs ensemble determined by these $[q]= \lbrace q_m \rbrace_m$, and the hydrodynamics equation is \eqref{conservation} under this constitutive relation or ``equation of state". 
This is simplified when considering linear response fluctuations over a (generalized) equilibrium state: to leading order, the current fluctuations are directly related to charge fluctuations $\delta j_n = A^{i}_n \delta q_i +\dots$ where the matrix $A^{i}_n = A^{i}_n[q]= \partial \j_n[q]/\partial q_i$ is a property of the underlying equilibrium state and the sum is over a basis of all conserved quantities---the matrix $A^{i}_n$ is entirely determined by thermodynamics (here and below, we use Einstein's convention of summing over repeated indices). The charge fluctuations then obey the equation $\partial_t \delta q_n +A^{i}_n \partial_x \delta q_i =0$, which defines the linearized Euler-scale hydrodynamics of the system. These linearized hydrodynamic equations can be solved by diagonalizing the matrix $A$: its eigenvectors define normal modes, propagating ballistically with effective velocities given by the eigenvalues $v_n^{\rm eff}$ of $A$. General principles of statistical mechanics, including the decay of correlations at large distances, guarantee that these eigenvalues are real; thus Euler-scale equations are hyperbolic.

The above general recipe can be formally applied to integrable systems. However, as there are infinitely-many linearly independent conserved quantities, and evaluating average currents is not part of the legacy integrability techniques, chances of success may {\em a priori} look rather dim. Happily, an alternative, defining feature of integrability reveals its underlying fundamental physics and allows for an elegant solution: it is the existence of infinitely long-lived ``quasiparticle excitations", even in high-temperature and entropy states. The expectation value of conserved charges can be expressed in terms of the density of such quasiparticles $q_n(x,t) = \int d\lambda\, q_n(\lambda) \rho_\lambda (x,t)$ where $\lambda$ is a parameter called alternatively {\em spectral parameter} or {\em rapidity} (we will use the latter), which parameterizes the quasiparticles' dispersion relation. The quantity $\rho_\lambda$ is the density of quasiparticles at rapidity $\lambda$, per unit length and rapidity, and $q_n(\lambda)$ corresponds to the amount of charge $Q_n$ carried by a single quasiparticle $\lambda$. The key breakthrough of Refs~\cite{Doyon, Fagotti} leading to the advent of GHD was the realization that the currents can also be expressed in a similar way at Euler scale (ignoring gradient corrections): $j_n(x,t) = \int d\lambda \,q_n(\lambda) v^{\rm eff}_\lambda [\rho(x,t)] \rho_\lambda (x,t)$. Here the effective velocity $v^{\rm eff}_\lambda [\rho(x,t)] $ is a function of the local generalized equilibrium state, thus a functional of the state function $\lambda\mapsto \rho_\lambda(x,t)$, and corresponds precisely to the eigenvalues of the matrix $A$ introduced above. Inserting these relations into the continuity equations, this yields 
\begin{equation} \label{eqGHDEuler}
\partial_t \rho_\lambda + \partial_x \left( v^{\rm eff}_\lambda [\rho] \rho_\lambda \right) =0.
\end{equation}
These equations for the quasiparticle densities $\rho_\lambda $ form the basis of Euler-scale GHD: they quantitatively describe ballistic transport in integrable systems. In terms of the structure of hydrodynamics, it turns out that one can find the Riemann invariants of the conservation law~\eqref{eqGHDEuler}, and write it in transport form, where $v^{\rm eff}_\lambda$ is the hydrodynamic velocity for the normal mode $\lambda$~\cite{Doyon, Fagotti}. Thus GHD displays a continuum of hydrodynamic normal modes, this being one of its main characteristics.

The effective velocity in eq.~\eqref{eqGHDEuler} has an appealing semi-classical interpretation in terms of a soliton gas~\cite{el2005,solitongases,BBH}. Quasiparticle excitations with rapidity $\lambda$ behave as semiclassical solitons and move with ``bare'' velocity $v^0_\lambda$ in the vacuum. In integrable systems, quasiparticles do not decay, and they only scatter ellastically with picking up rapidity-dependent phase shifts ${\rm e}^{2 i \delta}$, known from Bethe ansatz~\cite{Takahashi}. At a semi-classical level, these phase shifts induce Wigner time delays $\Delta \tau = 2 \hbar \frac{d \delta}{ d E}$, 
with $E$ the quasiparticle's energy~\cite{PhysRev.72.29,PhysRev.98.145}. These accumulated time delays due to scattering with other quasiparticles renormalize the velocity as follows: over a time $t$, a tagged quasiparticle with rapidity $\lambda$ moves over a distance $v^{\rm eff}_\lambda t$ instead of $v^{0}_\lambda t$ given by (Fig.~\ref{fig:hydro}b)
\begin{equation} \label{eqVeff}
v^{\rm eff}_\lambda = v^{0}_\lambda + \int d \mu \,\rho_\mu ( v^{\rm eff}_\mu - v^{\rm eff}_\lambda) \Delta x (\lambda, \mu).  
\end{equation}
Here, $\rho_\mu | v^{\rm eff}_\mu - v^{\rm eff}_\lambda| t$ is the (mean) number of quasiparticles with rapidity $\mu$ that will collide with the tagged quasiparticle $\lambda$, and  $\Delta x (\lambda, \mu) = v^0_\lambda \Delta \tau (\lambda, \mu)   $ is the resulting displacement due to the collision (where a positive sign means a recoil of the particle in focus, from the direction where the particle against which it collided came from). The effective velocity $v^{\rm eff}_\lambda$ that solves~\eqref{eqVeff} is a functional of the local generalized equilibrium state $\rho$ and depends non-trivially on interactions.

Smoothly varying space and time-dependent potentials can also be incorporated into the GHD framework~\cite{SciPostPhys.2.2.014,2019arXiv190601654B}. We briefly discuss the case of a spatially varying potential (e.g., an optical trap). The potential in each hydrodynamic cell can be treated as constant (i.e., a local density approximation can be made) when computing thermodynamics.
The spatial gradients of the potential induce forces, which can be treated analogous to the semiclassical theory for electron transport~\cite{ashcroft}: a potential $V(x)$ induces a force $F(x) = \partial_x V(x)$. This force accelerates each quasiparticle, causing its rapidity to evolve in time. This effect is incorporated in Eq.~\eqref{eqGHDEuler} by adding terms proportional to $F(x) \partial_\lambda \rho_\lambda(x)$. Deriving the correct coefficients is subtle because of backflow effects; we refer to Refs.~\cite{SciPostPhys.2.2.014,2019arXiv190601654B,PhysRevLett.125.240604,Durnin_2021} for details. There are also recent works considering effects of localized impurities \cite{rylands2023transport,hubner2023mesoscopic}

Our logic so far followed the same hydrodynamic principles as for any other Euler-scale equation, but leads to some peculiar consequences. 
In many contexts---such as the Riemann problem or just the dynamics from smooth inhomogeneous initial conditions---typical hyperbolic equations develop shocks~\cite{Bressan2013, olver2014introduction}, but GHD does not~\cite{Doyon, Fagotti,PhysRevLett.119.195301}. These effects arise because GHD has the property of {\em linear degeneracy}~\cite{rozdestvenskii1967imposs,liu1979development,el2011kinetic,DOYON2018570,BBH0,Bulchandani_2017}. In physical terms, this means that the hydrodynamic modes are not ``self-interacting". In addition to the presence of a continuum of hydrodynamic modes, the lack of self-interaction is at the basis of many of the special features of GHD.

The GHD equations~\eqref{eqVeff} can be solved efficiently numerically (and sometimes analytically \cite{PhysRevB.97.081111}) both at zero and finite temperature~\cite{VKM,PhysRevLett.119.195301,2019arXiv190601654B}, with an open source package available implementing most standard integrable models of interest~\cite{10.21468/SciPostPhys.8.3.041, MOLLER2023112431}. It has integrability structures~\cite{DOYON2018570,Bulchandani_2017}, and there are analytical solutions by integral equations where time appears explicitly~\cite{DOYON2018570,Bulchandani_2017} (paralleling the hodograph method~\cite{Tsarev}). The GHD equation first appeared within the context of soliton gases and hard rods~\cite{zakharov1971kinetic,Boldrighini1983,EL2003374}, where it was derived by different methods; in particular see the reviews \cite{El_2021,suret2023soliton} on soliton gases. The semiclassical picture recalled above was recently explained by seeing long-wavelength quantum many-body states as gases of interacting, sharp wave packets~\cite{doyon2023ab}.

\subsection{Diffusive corrections}
\label{refDiffusiveCorrections}
As discussed above, in hydrodynamics one looks for a long-wavelength expansion by projecting onto slow variables. Naturally, beyond the Euler scale, one cannot simply assume that a fluid cell is described by a (generalized) Gibbs ensemble: one must account for the partial relaxation of current fluctuations leading to the (G)GE, missed by Euler-scale GHD.
This effect can be obtained from higher-order terms in the hydrodynamic expansion~\cite{dbd1, ghkv, dbd2,GV19,10.21468/SciPostPhys.9.5.075,doyon2019diffusion,DeNardisReview_2023}. One projects onto spatial variations of conserved densities: $j_n = \j_n[q] - \mathfrak D_n^i[q]\partial_x q_i$ for some diffusion matrix $\mathfrak D_n^i$, which, by standard arguments, may be evaluated in terms of current-current correlations by the Kubo formula. Combining with advanced technology of integrable systems and some educated guesses, this gives the universal diffusion matrix of GHD \cite{dbd1,dbd2}.

There is another way of understanding what is going on. In linear response, one expands the currents about their equilibrium values, $\delta j_n = A^i_n \delta q_i - \mathfrak D_n^i\partial_x \delta q_i + \ldots$. By the conservation laws, this gives rise to simple linear diffusive equations for $\delta q_i$'s. But strong spatial variations also occur from non-linear deformations, whose time evolution will show diffusive relaxation.
%
%
%
It turns outs that one may use this effect to connect the non-linear current expansion with its diffusive expansion: it suffices to include terms that are quadratic in charge fluctuations~\cite{10.21468/SciPostPhys.9.5.075}, i.e., $\delta j_n = A^i_n \delta q_i + \frac{1}{2}R^{ij}_n \delta q_i \delta q_j + \ldots$, where
%
$R^{ij}_n = \frac{\delta^2}{\delta q_i \delta q_j} j_n$.
%
Intuitively, this corresponds to replacing $A^i_n$ with $A^i_n + R^{ij}_n \delta q_j/2$, i.e., incorporating the dependence on each quasiparticle velocity of the fluctuating densities of all other quasiparticles. One can use these expressions to explicitly compute the exact diffusion matrix of GHD \cite{10.21468/SciPostPhys.9.5.075}, and it turns out that this is indeed equivalent to the result obtained by using the Kubo formula thanks to a precise projection mechanism \cite{doyon2019diffusion}.
%
%
The resulting physics is simple to explain from a kinetic perspective~\cite{ghkv}, based on the soliton-gas interpretation of GHD~\cite{solitongases} and giving the same results in a transparent way. In an integrable system, each quasiparticle moves along a ballistic trajectory, which is either delayed or advanced by collisions. Collisions occur whenever the test quasiparticle encounters another quasiparticle, but this number experiences equilibrium thermal fluctuations. A quasiparticle traveling for a time $t$ passes through a region of size $L \sim t$, so the number of collisions it experiences fluctuates by $\sqrt{t}$. Therefore, the distance it travels fluctuates by $\sqrt{t}$. The quasiparticle's trajectory (which is a straight line at Euler scale) undergoes gaussian broadening as a result of collisions; this broadening is due to thermal fluctuations and increases with temperature.

We briefly contextualize this picture of diffusion in GHD and contrast it with the physics of diffusion in nonintegrable models. 
The most obvious difference is that diffusion in GHD arises as a \emph{subleading} correction to Euler-scale hydrodynamics, whereas in nonintegrable lattice models diffusion is the \emph{leading} effect, since the current has no overlap with any conserved charge density.
Moreover, the physical origin of diffusion in the two cases is different: Diffusion in GHD is purely due to their projection onto ``bilinear charges"---products of conserved quantities---representing the covariance of density fluctuations, while in non-integrable systems diffusion is due to the ``fast'' relaxation of the current. A consequence of this distinction, to which we will return in Sec.~\ref{secFCS}, is that the noise term in fluctuating GHD is of a different nature than that expected in non-integrable models~\cite{ferrari2023macroscopic}. Indeed, noise in fluctuating GHD is strongly correlated on diffusive distance scales, while noise in non-integrable models is only expected to be correlated on microscopic scales. A noise realization on a tagged quasi-particle comes from a train of other quasi-particles going through it, and this train---this realization---propagates with little changes on diffusive scales. The suggestion is that this is a general phenomenon of integrability.

Finally, even for nonintegrable systems that exhibit ballistic transport (e.g., a chaotic Galilean fluid), the nature of the ``diffusive-scale'' corrections is different than what we have described here. The most striking difference is that the broadening of a ballistic front is not asymptotically diffusive at all in chaotic systems: instead, it is superdiffusive, for instance the sound front having dynamical exponent $z = 3/2$ and scaling functions in the Kardar-Parisi-Zhang (KPZ) universality class \cite{spohn_nlfhd}. This result can be understood in terms of nonlinear fluctuating hydrodynamics~\cite{spohn_nlfhd,Spohn2014} and also from direct analysis of the Kubo formula~\cite{doyon2019diffusion}. The KPZ nonlinearity is due essentially to the self-interactions of the Euler-scale modes and is therefore absent in integrable systems.

In addition to these diffusive corrections, higher-order gradient corrections have also been explored~\cite{PhysRevB.96.220302,DeNardisReview_2023}, as well as quantum corrections~\cite{PhysRevB.96.220302,ruggiero2019quantum}, which we will not discuss more here.

\section{Experimental toolbox}

In this section we review some of the experimental tools that have been used to study transport and nonequilibrium dynamics in approximately integrable systems. Our focus will be on \emph{observables}: the past decade has seen a vast expansion in the scope of what can be measured; these newly accessible observables allow for stringent tests of theories like GHD and stimulate new theoretical questions (to which we will return in Sec.~\ref{secOpenQuestions}). First, however, we provide a very brief overview of how approximately integrable systems can be realized and driven out of equilibrium in present-day experiments. 

\subsection{Realizations}

\emph{Ultracold atoms}.---Much recent progress in studying integrable dynamics has involved cold-atom experiments. Cold atomic gases can be trapped in quasi-one-dimensional geometries by one of two methods: either one can apply a deep optical lattice potential along two spatial directions (ensemble of decoupled tubes) or one can directly trap individual gases using an atom chip. For details on these see Refs.~\cite{PhysRevLett.81.3108, RevModPhys.78.179, bloch_review, gross2017quantum,Folman2002a,reichel2011atom}. The one-dimensional gases are described by the Lieb-Liniger model~\cite{lieb1963, PhysRev.130.1616}, which assumes strictly contact interactions; for ultracold atoms this is an excellent approximation, because the typical interparticle spacing ($> 100$ nm) are much larger then the range of the van der Waals interaction ($\leq 1$ nm).
The two approaches are complementary: atom chips trap single 1D systems and allow to study all the details through correlations~\cite{Schweigler2017}, full distribution functions~\cite{nphys941}  and  matter-wave interference~\cite{Schumm2005, RevModPhys.81.1051} (see Sec.~\ref{intapp}), while the optical lattices typically allow for tighter confinement and stronger interactions~\cite{bloch_tonksgirardeau, Kinoshita20082004, doi:10.1126/science.1175850} (e.g., through the use of Feshbach resonances~\cite{Inouye1998, PhysRevLett.81.938, PhysRevLett.91.163201}) but suffer from ensemble averaging. 

In addition to the continuum Lieb-Liniger model, ultracold gases can be used to realize integrable lattice models such as the Hubbard and Heisenberg models. The essential modification is to add a third (weaker) optical lattice potential along the tubes where the atoms are confined. The motion along the tubes is described (in the appropriate parameter regime) by a Fermi- or Bose-Hubbard model, depending on the statistics of the underlying particles. The Fermi-Hubbard model is itself integrable~\cite{esslerBook}, and exhibits anomalous hydrodynamics~\cite{PhysRevB.102.115121,moca2023kardarparisizhang}. The Bose-Hubbard model is not integrable; however, for half filling and strongly interacting particles, charge fluctuations are energetically suppressed and the Bose-Hubbard model reduces to the integrable Heisenberg model. The symmetry of the spin-spin interactions in the resulting Heisenberg model depends on the microscopic state-dependent scattering properties of the atoms: in general it is approximately isotropic, but by tuning the system close to a state-dependent Feshbach resonance one can realize anisotropic Heisenberg models with tunable anisotropy~\cite{Jepsen:2020aa}. 

\emph{Solid-state magnets}.---In parallel with these advances in ultracold atomic experiments, there has also been a large body of work exploring spin and heat transport in quasi-one-dimensional magnets in the solid state. In particular, experiments on heat transport in the materials CaCu$_2$O$_3$ and SrCuO$_2$ have shown large contributions that have been attributed to long-lived magnons~\cite{HESS20191}. 
The spin dynamics in these materials is believed to be governed by the one-dimensional antiferromagnetic Heisenberg model. An especially dramatic manifestation of integrable dynamics was recently discovered in the one-dimensional Heisenberg magnet, KCuF$_3$, in which neutron scattering experiments showed clear evidence of superdiffusive spin transport at temperatures exceeding $100$ K~\cite{2020arXiv200913535S}. 
A complementary solid-state platform uses Hamiltonian-engineering methods on \emph{nuclear} spins in the material fluorapatite~\cite{wei2019emergent}. In fluorapatite, both the isotropic and anisotropic Heisenberg spin chain can be realized; a key prediction of GHD, namely that the easy-axis XXZ spin chain exhibits ballistic energy transport but diffusive spin transport, was recently verified in this experimental system~\cite{paipeng}.

\emph{Qubit devices}.---In recent years, intermediate-scale quantum computers have become powerful tools for quantum simulation. These systems are appealing because they allow one to probe a wealth of otherwise inaccessible observables. Unlike the systems we described above, these devices implement \emph{gate-based}, discrete-time dynamics. To approximate Hamiltonian evolution, one has to perform a Trotter decomposition on the Hamiltonian. In general Trotter errors to lead to integrability breaking, but remarkably the XXZ spin chain has a family of discrete-time deformations that remain exactly integrable~\cite{PhysRevLett.121.030606}. Quantum simulations of these discrete-time integrable systems have become an  widely used approach to understand their transport properties~\cite{GooldKPZ,rosenberg2023dynamics}. 

\subsection{Transport}

One of the most natural probes of GHD is linear-response transport. Since the dynamics at the very latest times is dominated by integrability-breaking effects, to probe the integrable regime one must work at finite frequency. The finite-frequency, finite-momentum density response is controlled by the dynamical response function $C(x,t) = \langle n(x,t) n(0,0) \rangle$, where $n(x,t)$ is the density at position $x$ and time $t$, and the expectation value is taken in a thermal equilibrium state. (For spin models like the Heisenberg model, we can define $n(x) = \frac{1}{2}(1 + \sigma^z_x)$.) Equivalently, one can consider the dynamical structure $S(q,\omega)$, which is the Fourier transform of $C(x,t)$. 

The density response function can be extracted in any of the realizations we have mentioned. In solid-state magnets, $S(q,\omega)$ has been measured using neutron scattering, while in ultracold gases, it has been measured using Bragg scattering~\cite{PhysRevLett.115.085301} for the Lieb-Liniger model. In superconducting qubits and 1D optical lattices, it is more convenient to work in real space: both platforms have single-site imaging capabilities. While neither platform allows direct access to multitime correlations, there are two standard real-space approaches, the typicality approach and the weak quench approach, for extracting these correlations. We now briefly review both. 

The typicality approach is simplest to describe at infinite temperature. The infinite-temperature density matrix can be written as $\rho = D^{-1} \sum_{i = 1}^D \ket{i}\bra{i}$, where $D = 2^L$ is the dimension of Hilbert space and $\ket{i}$, the binary representation of $i$, is an eigenstate of $\sigma^z_i$ at every site $i$. In terms of this decomposition, we can write $C(x,t) = D^{-1} \sum_{i = 1}^D \langle i | n(x,t) n(0,0) | i\rangle$. Note that $\ket{i}$ is an eigenstate of $n(0,0)$. This allows us to estimate $C(x,t)$ as follows: we begin with a random product initial state, record the density at site $0$, time evolve, measure $\langle i | n(x,t) | i \rangle$, and multiply this result by the initial density at site $0$ to get one sample of $C(x,t)$. Then we repeat this procedure many times and average to get an estimate of $C(x,t)$. The typicality approach is experimentally straightforward for any density matrix that can be decomposed in terms of product states; it has also been used numerically with Krylov-space evolution~\cite{PhysRevB.80.184402,HeitmannRichterSchubertSteinigeweg}.

In the weak quench approach, one instead initializes the left and right halves of the system in equilibrium states at slightly different chemical potentials, i.e., $\rho_\mu = \rho_0 \exp(\mu \sum\nolimits_x \mathrm{sgn}(x) n(x))$. Taylor expanding to leading order in $\mu$, one finds that $C(x,t)$ is the spatial derivative of the expectation value $\mathrm{Tr}(n(x,t) \rho_\mu)$ in the limit $\mu \to 0$. Therefore, once again, one can use a single-time expectation value to extract the structure factor. This method is much less demanding than the typicality approach in terms of state preparation, but it has an inherent tradeoff between signal to noise ratio (worse at small $\mu$) and the validity of linear response.

\begin{figure}[t!]
	\centering 
	\includegraphics[width=.99\linewidth]{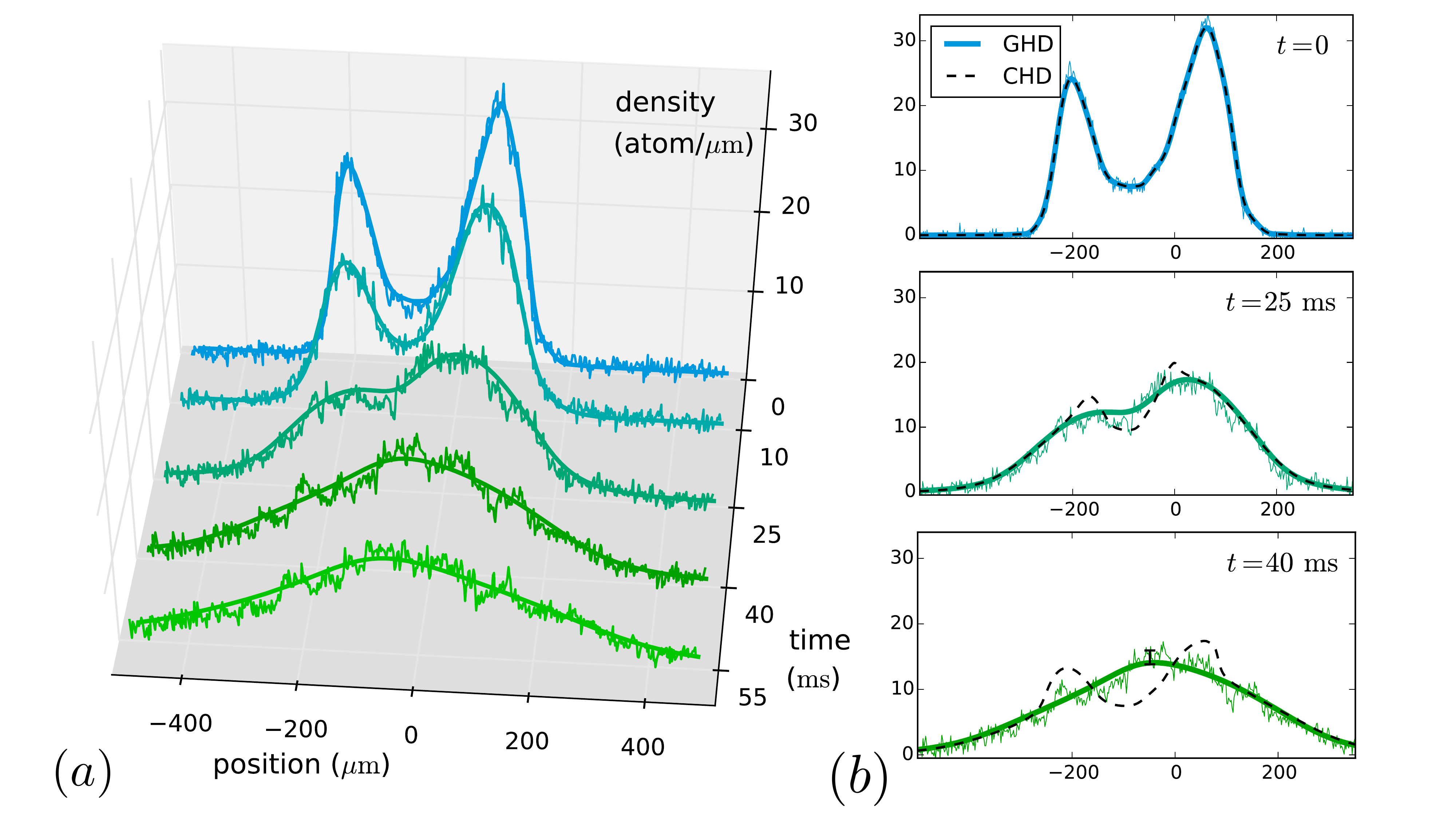}
 
	\caption{{\bf Density dynamics.} (a) Density dynamics of an expanding 1D Bose gas initially trapped in a double-well potential, compared with GHD. (b) Comparison of the predictions of GHD and conventional hydrodynamics (CHD) including only momentum, energy and particle number: taking into account higher-order conservation laws is clearly crucial to capture quantitatively the dynamics. Reproduced from Ref.~\onlinecite{PhysRevLett.122.090601}.}
	\label{fig:densitydynamics}
\end{figure}

\subsection{Density dynamics and full counting statistics}\label{intdensityFCS}

GHD is not limited to near-equilibrium transport, but makes quantitative predictions for the coarse-grained dynamics of densities from nonequilibrium initial states. Cold-atom experiments that are capable of in-situ imaging can very naturally access these dynamics. This technique was used for the first direct experimental test of GHD on an Atom Chip setup~\cite{PhysRevLett.122.090601}, where the density evolution of a single, weakly interacting quasicondensate following quenches of the 1D potential was studied.
In one quench, atoms in thermal equilibrium were initially confined in a double-well potential and then let to expand freely in 1D (Fig.~\ref{fig:densitydynamics}).
The subsequent evolution of the atomic density demonstrated excellent agreement with GHD predictions, whereas conventional hydrodynamics based on the assumption of local thermal equilibrium failed~\cite{PhysRevA.89.013621, PhysRevLett.119.195301}.
In another measurement, akin to the quantum Newton’s cradle, the double-well was quenched to a harmonic potential.
Again, GHD predictions correctly captured the main observed features, however, at longer evolution times the dynamics relaxed faster than predicted. (We will return to this discrepancy in Sec.~\ref{secIntBreaking}, where we discuss integrability-breaking.)

Continuum experiments with in-situ imaging usually measure coarse-grained densities, allowing one to extract the dynamics of expectation values (which is the standard use case of hydrodynamics). However, optical lattice experiments using quantum gas microscopy~\cite{Bakr2009, Sherson2010} (as well as experiments with superconducting qubits) can take simultaneous snapshots of all the atoms in a system. The statistics of these snapshots encode, not just expectation values, but fluctuations of arbitrarily high order. The study of these fluctuations, termed ``full counting statistics'', is an active area of theoretical exploration, to which we will turn in Sec.~\ref{secFCS}. As was realized very early on, atom-chip experiments also allow for interferometric measurements of fluctuation statistics, but the quantity whose fluctuations are measured in that setup is the \emph{superfluid phase} rather than the density (Sec.~\ref{intapp}).

In superconducting qubit systems (and to some extent in quantum gas microscopes), one can also explore how the expectation values of more complicated conserved quantities (such as the energy) evolve~\cite{Maruyoshi_2023,googleboundstates}. This can be done by separately measuring all the Pauli strings that appear in the energy, and combining their expectation values. Measuring the fluctuations of these higher-order conserved quantities is more challenging, and has not yet been accomplished.

\subsection{Direct rapidity measurements}

\begin{figure}[t!]
	\centering 
	\includegraphics[width=.99\linewidth]{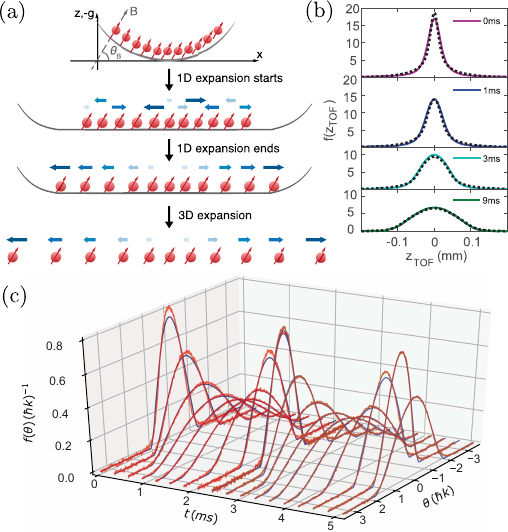}
    \phantomsubfloat{\label{fig:rapidity_distibution_a}}
    \phantomsubfloat{\label{fig:rapidity_distibution_b}}
    \phantomsubfloat{\label{fig:rapidity_distibution_c}}
    \vspace{-2\baselineskip}
    
	\caption{{\bf Rapidity distribution measurement and dynamics.} (a) Schematic illustrating the modified time-of-flight protocol for measuring the rapidity distribution of a 1D gas. First, the confining longitudinal potential is removed, allowing the atoms to expand in 1D. Next, the transverse trapping is removed, resulting in 3D expansion followed by time-of-flight absorption imaging. Reproduced from Ref.~\onlinecite{PhysRevA.107.L061302}. (b) Density profiles measured after different durations of 1D expansion (solid lines) compared with predictions for hard-core bosons (dotted lines). For sufficiently long expansion times, momentum distribution converges towards the rapidity distribution. From~\cite{wilson2019observation}. (c) Evolution of rapidity distribution following a sudden increase of the depth of a Gaussian longitudinal confinement. The experimental observations (red curves) are compared to GHD predictions for a gas initially in the ground state (blue curves). Reproduced from Ref.~\onlinecite{2020arXiv200906651M}.}
	\label{fig:rapidity_distribution}
\end{figure}

GHD is, at its core, a theory of how the full quasiparticle distribution evolves. 
A major advance in the experimental testability of GHD was the modified time-of-flight protocol of Ref.~\cite{wilson2019observation}, which allows for direct extraction of rapidity distributions under mild assumptions (Fig.~\ref{fig:rapidity_distibution_a}). 

In time-of-flight measurements of the momentum distribution, the confining potential along all directions is suddenly turned off and the gas is let to freely expand. 
Following sufficiently long expansion times $t$, the in-trap momentum distribution $n(p)$ maps to the atomic density $n(x = p t/m)$, which can then be measured by standard imaging techniques~\cite{PhysRevA.83.031604}.
In the 1D Bose gas, atomic momenta are generally complicated superpositions of rapidity states, while rapidities have the physical interpretation of \textit{asymptotic} scattering momenta~\cite{PhysRevLett.114.125302}.
Thus, their distribution can be extracted following a \textit{one-dimensional} expansion of the gas: By switching off the confining potential only along the 1D tube, the system still evolves under the integrable Hamiltonian, whereby the rapidities of quasiparticles are conserved.
As the quasiparticles fly apart from one another, the atomic density $n$ within each tube becomes very dilute, causing the system to enter the strongly ``fermionized'' regime $\gamma = c/n \gg 1$, in which the system can be regarded as a noninteracting Fermi gas~\cite{rigol_muramatsu}, 
for which momentum and rapidity distributions coincide~\cite{RevModPhys.83.1405}.
Thus, measuring the momentum distribution following an initial 1D expansion yields the in-trap rapidity distribution (Fig.~\ref{fig:rapidity_distibution_b}). Similar ideas were previously used to extract rapidities from numerical simulations~\cite{PhysRevLett.80.3678, rigol_muramatsu, PhysRevLett.114.125302}.


Employing direct rapidity measurements, more stringent tests of GHD were possible: In 
Ref.~\cite{doi:10.1126/science.abf0147}, an ensemble of strongly interacting 1D gases was subjected to a drastic quench of the (quasi-)harmonic longitudinal potential, compressing the gas and inducing a breathing motion in good agreement with zero-entropy GHD (Fig.~\ref{fig:rapidity_distibution_c}).
By measuring both the momentum and rapidity distributions, the evolution of both kinetic and interaction energy of the system was monitored: 
In the hard-core regime, overlap between atoms is negligible, whereby the interaction energy vanishes.
However, at its most compressed state the gas leaves the hard-core regime, which was observed as a considerable dip in kinetic energy, again in agreement with GHD predictions.

In principle the rapidity measurements can be extended (in conjunction with state-sensitive imaging and band-mapping techniques) to lattice models such as the Hubbard model in which the vacuum is an empty state with no particles.
In such systems, one can generally implement a one-dimensional expansion by simply turning off the confining potential along the relevant direction.
Conceptually one could also implement this method for spin systems like the XXZ spin chain by surrounding the system with a large ``buffer'' of spin-polarized atoms; however, in practice there are many challenges, such as (e.g.) the necessity of an unreasonably large buffer to get good time-of-flight images, and this has yet to be demonstrated experimentally.

\subsection{Interferometric approaches}\label{intapp}

Matter-wave interferometry is a powerful tomographic technique that sidesteps some of the enormous complexity~\cite{Flammia_2012} contained in microscopic degrees of freedom, and focuses directly on emergent variables such as the matter-wave phase.
%
%
%
Starting from two parallel 1D gases, e.g., trapped on an atom chip, an abrupt removal of the trapping potential leads to a rapid transverse expansion of the two condensates, causing them to overlap in space and interfere~\cite{Schumm2005, RevModPhys.81.1051}.
From the measured interference pattern  several quantities used to characterize the system can be obtained:
Full distribution functions (FDFs) of the interference contrast are a direct measure of all even  correlations of the relative phase between the gases and hence determines the state of the system in detail~\cite{FCSGritsev, Hofferberth:2007aa, gring, Adu_Smith_2013, PhysRevLett.110.090405}.
FDFs have been used to demonstrate the emergence of a prethermalized state following a homogeneous quench of the system (see Sec.~\ref{sec:local_relaxation}).
Alternatively, the relative phase and correlation functions thereof can be extracted directly from the interference patterns~\cite{gritsev_demler}:
By monitoring the evolution of two-point correlations, a light-cone-like spreading of prethermal correlations was found~\cite{Langen2013}.
Later, higher order correlations were used to demonstrate the emergence of a GGE~\cite{Langen207}, while the connected correlation functions revealed the realization of a  gapped, strongly correlated effective field theory, namely the integrable sine-Gordon model~\cite{Schweigler2017,PhysRevX.10.011020,Schweigler2021}.

In thermal equilibrium, one-particle irreducible (1PI) correlation can be extracted revealing the momentum dependent quantum corrections to the Hamiltonian, the propagators and the change of the coupling parameters with momentum ('running couplings') giving deep inside into the emerging model describing the coarse grained physics ~\cite{PhysRevX.10.011020}.
A detailed measurement of the fluctuations in the phase and density of the 1d systems reveal a path to strong spin squeezing (close to 10 dB) and even more surprisingly non equilibrium dynamics of quantum properties ('squeezing oscillations') in a regime when the global mean field parameters are stationary~\cite{zhang2023squeezing}. 
Finally many-body tomography~\cite{Gluza2020} allowed to ectract von Neumann entropy and verify the  area law of mutual information~\cite{Tajik2023}, a first step on the path towards probing entanglement in continuous 1D quantum systems.

\section{Recent developments and open questions}
\label{secOpenQuestions}
The preceding sections introduced the framework of GHD and the experimental toolbox that has been used to explore it. As we saw GHD makes quantitatively accurate predictions about the large-scale dynamics of conserved densities in any integrable system, based purely on knowledge of its quasiparticle content. The \emph{qualitative} picture that emerges from this framework is simple and general. GHD becomes applicable after some $O(1)$ timescale that it is natural to associate with two-body elastic collisions. In the GHD regime, charge is transported by ballistically propagating quasiparticles, and the trajectory of each quasiparticle broadens diffusively because of its elastic collisions with other quasiparticles. This qualitative picture suggests a natural extension to nonintegrable systems: integrability-breaking causes quasiparticles to backscatter at some rate, but the quasiparticle properties are otherwise unchanged. 

Recent work suggests that both integrable and near-integrable dynamics are much richer than this caricature might suggest: the dynamics of integrable systems can itself be anomalous, and the effects of integrability-breaking can be unexpected. These anomalies are tied to the anomalous nature of fluctuations in integrable systems. Characterizing high-order fluctuations in integrable systems leads to the observation that GHD is in fact capable of describing the dynamics of integrable systems far beyond simple expectation values: indeed, it can even capture the exponential decay of non-hydrodynamic correlation functions. In what follows we describe the current status of these ongoing developments and the associated open questions.



\subsection{Local relaxation and onset of hydrodynamics} \label{sec:local_relaxation}

GHD relies on local relaxation to a GGE.
%
When this GGE arises is therefore a central question (which lies outside GHD itself). 
Local relaxation was studied in a series of quench experiments 
in which a gas is initialized in a single 1D tube, which is then coherently split into two parallel tubes (i.e., the cross-section of the tube is morphed from a single well to a double well). 
The splitting procedure creates a nonequilibrium state~\cite{Hofferberth:2007aa}, whose subsequent evolution can be measured using matter-wave interferometry. This quench is spatially homogeneous, so the relaxation probed is in some sense explicitly local.

Measurements of the full distribution functions (FDFs) of the interference contrast revealed the appearance of a thermal-looking relaxed state~\cite{gring, Adu_Smith_2013}, although the system was clearly not in thermal equilibrium, as the corresponding temperature was much lower than the initial one.
Instead, the measured relaxation dynamics manifested prethermalization.
By monitoring the phase coherence between the two condensates, it was revealed that the thermal correlations of the prethermalized state emerge locally essentially immediately in their final form and then propagate through the system~\cite{Langen2013} in a light-cone-like fashion; prethermal correlations were established within distances proportional to the speed of sound, whereas the initial phase coherence produced by the splitting process was preserved at larger separations.
The evolution continued until a quasi-steady state was reached, characterized by exponentially decaying correlations throughout the entire system.

Analysis of two-point (and higher order) phase correlations after a series of different splitting quenches demonstrated that this prethermalized state corresponded to a GGE~\cite{Langen207}; the extracted occupations of different Bogoliubov modes could not be described using a thermal distribution at a single temperature, but instead required the multiple Lagrange multipliers of a GGE.
Further, the extracted higher order correlations exhibited clear distinction from those of a thermal state.
Following the initial relaxation to a GGE, a much slower relaxation towards thermal equilibrium was observed.
Although integrable systems generally do not thermalize by themselves, the process can be facilitated by a weak breaking of integrability -- see section~\ref{secIntBreaking} below.
In any case the phonon basis in which the system was measured does not correspond to the true eigenstates of the Lieb-Liniger model, leading to some dephasing via integrable dynamics~\cite{PhysRevLett.130.140401, PhysRevX.12.041032}.\\

Heavy ion collision studies~\cite{Ralf_Rapp_2004} suggest that hydrodynamics can occur on timescales even shorter than local equilibration, in a process dubbed hydrodynamisation~\cite{Florkowski_2018, Schenke_2021, RevModPhys.93.035003}. 
The hydrodynamisation of a Lieb-Liniger gas was studied in a recent experiment~\cite{hydrodynamization}:
Following a high energy quench in the form of a Bragg pulse sequence, rapid redistribution of energy among distant momentum modes was observed.
The atomic momenta are generally complicated superpositions of rapidity states, whose relative phase evolution changes the occupancy of momentum modes.
Hence, the relaxation time scale is set by the highest energy difference between rapidity states.
Meanwhile, the rapidity distribution, which also was measured, remains constant on these time scales.
Next, the observed hydrodynamisation was succeeded by local prethermalization, where the momentum distribution equilibrates through the spreading of excited quasiparticles. 
The time scale, noted to be surprisingly fast, was observed to vary inversely with momentum.

\subsection{Anomalous transport}


Perhaps the best-studied integrable system is the XXZ spin chain, 
\begin{equation}\label{xxzham}
H = -J \sum\nolimits_i (S^x_i S^x_{i+1} + S^y_i S^y_{i+1} + \Delta S^z_i S^z_{i+1}).
\end{equation}
In this model, $Z_{\mathrm{tot}} \equiv \sum_i S^z_i$ is conserved; we will call the transport of this conserved quantity spin transport. Spin transport is ballistic~\cite{PhysRevLett.82.1764, karraschdrude, Prosen20141177,PhysRevB.96.081118,PhysRevLett.111.057203,PhysRevLett.119.020602, urichuk2019spin}  (with a Drude weight predicted by GHD~\cite{PhysRevB.97.081111, PhysRevLett.119.020602, BBH, PhysRevB.96.081118}, with anomalous subleading corrections~\cite{2019arXiv190905263A}) when $|\Delta| < 1$, in the so-called easy-plane regime. When $|\Delta| > 1$, the easy-axis regime, spin transport is diffusive. Interestingly, at the critical point $\Delta = 1$---i.e., the isotropic Heisenberg spin chain---spin transport is superdiffusive, see Ref.~\onlinecite{Bulchandani_2021} for a recent review. 
The transport of energy and other charges that are even under spin flip remains ballistic for all $\Delta$. In fact, even spin transport is ballistic when the initial state explicitly breaks spin-flip symmetry. Thus, the absence of ballistic spin transport, and in particular the emergence of super-diffusion, is a consequence of the interplay between integrability and symmetry~\cite{PhysRevLett.119.020602}. 

\subsubsection{Easy-axis XXZ model}
\label{secXXZ}
We begin by analyzing the easy-axis regime of Eq.~\eqref{xxzham}, which illustrates many of the features of the isotropic point in a simpler context. To construct the quasiparticles in this regime, one starts with the ferromagnetic vacuum state and flips spins~\footnote{In the Bethe ansatz construction, the vacuum state is ferromagnetic regardless of the sign of the interaction.}. Because of the symmetry, the vacuum is doubly degenerate. Starting from the vacuum, in the large-$\Delta$ limit, one creates excitations by flipping contiguous blocks of $s$ spins. These quasiparticles are called $s$-strings. Because $Z_{\mathrm{tot}}$ is conserved, $s$-strings can neither grow nor shrink; because of energy conservation, they cannot break up. An $s$-string can, however, collectively move with a characteristic velocity $v_s \sim \Delta^{1-s}$. The $s$-strings are qualitatively the same for all $|\Delta| > 1$ but in general are perturbatively dressed. The integrability of the model guarantees that $s$-strings remain stable even at finite density. 

To understand the peculiar features of transport in this model, let us pick the all-$\downarrow$ vacuum and construct two quasiparticles above it, an $s = 1$ string (i.e., a magnon) and an $s \gg 1$ string (i.e., a large domain of $\uparrow$ spins). We now consider how these quasiparticles scatter off each other. The magnon moves ballistically as a minority $\uparrow$ spin while the domain is essentially frozen. When the magnon gets to the edge of the domain, it flips its polarity and travels through the domain as a minority $\downarrow$ spin. For spin to be conserved during this process, two $\uparrow$ spins are deposited at the interface when this happens. The process repeats in reverse when the magnon leaves the domain. Therefore, when a magnon passes through a domain, (i) the magnon is transiently demagnetized, and (ii)~the domain moves by two sites. 

We now extrapolate these observations to the equilibrium state, in which both magnons and domains are present at high density. Since (in the equilibrium state) the magnon passes through as many $\uparrow$ as $\downarrow$ domains, it carries no spin on average. Moreover, the passage of many magnons through a domain gives it random displacements and causes it to diffuse over time. At finite $\Delta$, of course, there is no sharp distinction between magnons and domains: rather, on a given timescale $t$, $s$-strings with $\Delta^{1-s} t \ll 1$ are diffusive and polarized while smaller strings are ballistic and depolarized. In the $t \to \infty$ limit, spin transport is due entirely to the $s \to \infty$ string, which has zero velocity and is strictly diffusive. In the $\Delta > 1$ regime the diffusion constant of this giant quasiparticle is set by its collisions with magnons, and thus remains finite as $t \to \infty$. This diffusion constant can be quantitatively computed~\cite{GV19, dbd2,KnapTracer}  using the ideas from Sec.~\ref{refDiffusiveCorrections}: it is simply the spread in the (zero-velocity) trajectory of the giant quasiparticle due to its collisions with all other quasiparticles. (Note that to carry out this calculation within GHD, one must compute the diffusion constant in a state that weakly breaks the $\mathbb{Z}_2$ symmetry with a small field $h$, and take the $h \to 0$ limit at the end of the calculation.) This picture in terms of ballistic quasiparticles giving rise to diffusive transport was already appreciated well before the advent of GHD~\cite{PhysRevLett.78.943, PhysRevLett.95.187201, PhysRevB.57.8307}.

The absence of ballistic transport can be directly linked to the $\mathbb{Z}_2$ symmetry of both the Hamiltonian and the initial state. In the quasiparticle picture, this symmetry is what leads light quasiparticles to depolarize \emph{exactly}. Within this picture, a global $\mathbb{Z}_2$ flip inside a region of size $\ell$ would just correspond to adding an $\ell$-string to the state; this would not change the state of the other quasiparticles, so quasiparticles at any finite $s$ are neutral under a global $\mathbb{Z}_2$ symmetry.

\subsubsection{Heisenberg model}

The Heisenberg model features the same quasiparticle hierarchy as the easy-axis XXZ model; however, since the underlying symmetry of the model is $SU(2)$ rather than $\mathbb{Z}_2$, the excitations are quadratically dispersing Goldstone modes rather than domains. The $s$-strings are solitonic wavepackets made up of Goldstone modes~\cite{PhysRevLett.125.070601}, such that each $s$-string (on top of the vacuum) carries spin $s$. The scattering among quasiparticles is qualitatively unchanged from the easy-axis regime: small quasiparticles are depolarized by passing through larger ones, and large quasiparticles undergo Brownian motion from their collisions with smaller ones. Crucially, however, large solitons are much more mobile than large domains: the group velocity of an $s$-string at the Heisenberg point scales as $v_s \sim 1/s$. Consequently the diffusion of giant quasiparticles is dominated by their collisions, not with magnons, but with other giant quasiparticles. 

The superdiffusive nature of transport~\cite{PhysRevLett.122.210602, lzp, idmp, GV19, NMKI19, vir2019,ye2022universal,PhysRevE.100.042116} can be seen from the following logic~\cite{PhysRevLett.127.057201}. The Kubo formula relates the diffusive term of the hydrodynamic expansion of the current, to the autocorrelation function of the current, which can be written as a sum of contributions from $s$-strings of all sizes. Each $s$-string contributes a current that is its charge ($s$) times its velocity ($\sim 1/s$), until it encounters a larger string. Straightforward thermodynamic constraints impose $\rho_s \sim 1/s^3$. Therefore, an $s$-string encounters a larger string on a timescale $\tau_s \sim s^3$. The autocorrelator of the current at time $t$ is given by the expression
\begin{equation}
\langle j(t) j(0)\rangle \sim \sum\nolimits_s \rho_s e^{-t/\tau_s} \sim \sum\nolimits_s s^{-3} e^{-t s^3} \sim t^{-2/3},
\end{equation}
from which it immediately follows that the a.c. conductivity goes as $\sigma(\omega) \sim \omega^{-1/3}$, giving superdiffusion with the dynamical exponent $z = 3/2$~\cite{GV19, NMKI19, PhysRevLett.125.070601,PhysRevLett.123.186601}. 

A version of the argument above can be made fully quantitative within the GHD framework, and correctly predicts not only the transport exponent but also the prefactor, $\sigma(\omega) = C \omega^{-1/3} + \ldots$. At present, no direct calculation exists for the dynamical structure factor, i.e., the $q$-dependent conductivity $\sigma(q,\omega)$. Numerical evidence strongly points to the dynamical structure factor following the KPZ scaling form~\cite{PhysRevLett.122.210602}, but understanding the origin of this scaling form remains an open question~\cite{vir2019,PhysRevLett.128.090604,denardis2023nonlinear,krajnik2023universal}. The superdiffusive nature of spin transport and associated KPZ exponents have also been observed  both in neutron scattering~\cite{Scheie2021} and quantum microscopy experiments~\cite{wei2022quantum} (Fig.~\ref{fig:FCS}a and b).


\subsection{Breaking integrability}

\label{secIntBreaking}

GHD is a kinetic theory of ballistically-moving quasiparticles whose velocity is dressed by interactions.
Transport properties can be
inferred from the fact that quasiparticles move ballistically and carry some charge (energy, magnetization {\it etc.}). Since integrability is never exactly realized in a realistic experimental setup, it is crucial to extend GHD to nearly-integrable systems~\cite{Bastianello_2021}. 
We now consider perturbing an integrable system with Hamiltonian $\hat{H}_0$ by a non-integrable perturbation $g \hat{V} = g\int dx\,\hat v(x)$ that destroys all but a few conservation laws.

Intuitively, the leading effect of the non-integrable perturbation is to thermalize quasiparticle distributions at long times: quasiparticles can scatter into one another, and acquire a finite lifetime. In other words, integrability-breaking endows the integrable Boltzmann dynamics with a collision term. 
%
Thus, generic hydrodynamic equations should arise, characteristic of non-integrable systems. If there are no residual currents that overlap with conserved quantities, as in typical non-integrable spin chains, then the result is purely diffusive hydrodynamic equations (for the energy, spin, etc.). Otherwise, there is remaining ballistic transport, and Euler-scale hyperbolic equations arise, generically with super-diffusive corrections (in 1D). We now discuss the transient from GHD to such hydrodynamic equations.

\subsubsection{Golden Rule approaches and their challenges}

The main effect of the integrability-breaking perturbation $g \hat{V}$ is to break most conservation laws in the system:
\begin{equation} \label{eqGHDchargebreaking}
\partial_t  q_n  + \partial_x j_n = {\cal I}_n \left[ q \right]. 
\end{equation}
Here the left-hand side of those equations corresponds to the ordinary GHD, where $j_n = \j_n[q] - \sum_i \mathfrak D_n^i[q]\partial_x q_i + \ldots$ is expressed as a gradient expansion of all the expectation value of the charges $[q] = \lbrace q_m \rbrace_m$. Meanwhile, the right-hand side ${\cal I}_n$ is due to integrability breaking, and will spoil the conservation of most charges. The quantity ${\cal I}_n$ is a generalization of the Boltzmann collision term~\cite{friedman2019diffusive,0305-4470-25-14-020}, and indeed when applying the general formalism to free-particle models (which are examples of integrable models), it is the Boltzmann collision term plus its higher-particle scattering generalization.

Some of the qualitative effects of the integrability-breaking term are relatively straightforward: most charges $Q_n = \int dx\, q_n$ now decay as functions of time on a characteristic time scale set by ${\cal I}_n$. In terms of a transport of the remaining conserved charges, assuming that only energy is conserved, this generically leads to a broadening of Drude peaks into Lorentzians, giving rise to additional diffusive effects in transport~\cite{friedman2019diffusive,2020arXiv200411030D,Bastianello_2021}, in a simple generalization of the Drude model. By contrast to the un-broken GHD equation, this is diffusion of standard type, associated to noise due to discarded small-scale degrees of freedom (Sec.~\ref{refDiffusiveCorrections}). 

However, other qualitative effects are slightly less clear. If momentum is preserved by the integrability-breaking term---for instance in Galilean-invariant systems---then the situation for transport is more complex, as, by our discussion above, super-diffusion should emerge. How this occurs is more nebulous; for instance, it is currently not understood how KPZ super-diffusion arises from breaking the integrability of a typical Galilean-invariant integrable system (for instance, from the Toda model~\cite{DoyonToda,Cao_2019Toda,Bulchandani_2019Toda} to a generic anharmonic chain).

Further, even with a general understanding of the qualitative features of the crossover from the integrable to the non-integrable phenomenology due to integrability breaking, a quantitative description of the dynamics is extremely challenging. This is largely due to the difficulty in evaluating the right-hand side  ${\cal I}_n$ in eq.~\eqref{eqGHDchargebreaking}. For weak perturbations $g \ll 1$, this term can be evaluated perturbatively using Fermi's Golden rule: 
\begin{equation} \label{eqFGRcharge}
{\cal I}_k   =4 \pi^2 g^2 \sum_n \delta (\Delta \varepsilon) \delta (\Delta p) \Delta q_k 
| \langle n| \hat{v} | \rho \rangle |^2,
\end{equation}
where $\Delta q_k $ is the difference of the eigenvalue of $Q_k$ on the state $| n \rangle$ compared to the  state $| \rho \rangle$ of the system (${\cal I}_k $ is a functional of $\rho$);
and similarly for $\Delta \varepsilon$ and $\Delta p$ with $\varepsilon$ the energy and $p$ the momentum. The delta functions in this expression restrict the sum over processes respecting energy and momentum conservation, since these are residual conserved quantity here as the perturbation is time-independent and homogeneous~\cite{bastianello2020generalised,2020arXiv200411030D} (in spin chains, the momentum is a compact variable; the formula needs to be interpreted accordingly, and momentum is therefore not a residual local or quasi-local  conserved quantity). It is possible to generalize this formula to time-dependent perturbations, and to write down the sum over accessible states $\sum_n$ in terms of general particle-hole excitations over the background state  $| \rho \rangle$. However, this general Fermi-Golden rule approach~\cite{friedman2019diffusive,2020arXiv200411030D,Bastianello_2021} faces two major difficulties: (1) the matrix elements (form factors) $\left| \langle n| \hat{v} | \rho \rangle \right|$ of the perturbation are unfortunately not known in general, and (2) even if they were, carrying out the sum over accessible states would be extremely challenging. This program can be carried out explicitly in the case of smooth noise perturbations: this restricts the accessible phase space to single particle-hole excitations with low momentum transfer, for which matrix elements can be evaluated explicitly within GHD~\cite{bastianello2020generalised}. Various approximations have also been explored in the literature~\cite{PhysRevB.103.L060302,PhysRevLett.126.090602}. 
However, addressing systematically generic integrability-breaking perturbations remains a very important challenge for future works.

External space-dependent potentials also generically break integrability. This is not seen at the Euler scale, but a calculation of diffusion in the presence of external potentials~\cite{PhysRevLett.125.240604} show integrability breaking effects of a type similar to the ``generalized Boltzmann" form \eqref{eqFGRcharge}. The dynamics within such an external potential and the process for long-time thermalization of integrable systems is however more complicated, including an apparently chaotic regime of time coming from the formation of small structures in phase space, before a slow, diffusion-led thermalization~\cite{PhysRevLett.120.164101,PhysRevLett.125.240604,biagetti2023three}. This reveals  turbulent-like effects~\cite{biagetti2023three} which are still largely to be investigated.

We also note that several works have hinted at broad regimes of anomalous (non-diffusive) transport in nearly-integrable systems~\cite{PhysRevLett.127.057201,2021arXiv210913251D}, where some quasi-particles remain very long-lived even in the presence of integrability-breaking perturbations. These lead to transport anomalies, apparently of a different nature than those seen in non-integrable Galilean systems. Much still needs to be understood in this direction.

\subsubsection{Dimensional crossover}

\begin{figure}[t!]
	\centering 
	\includegraphics[width=.99\linewidth]{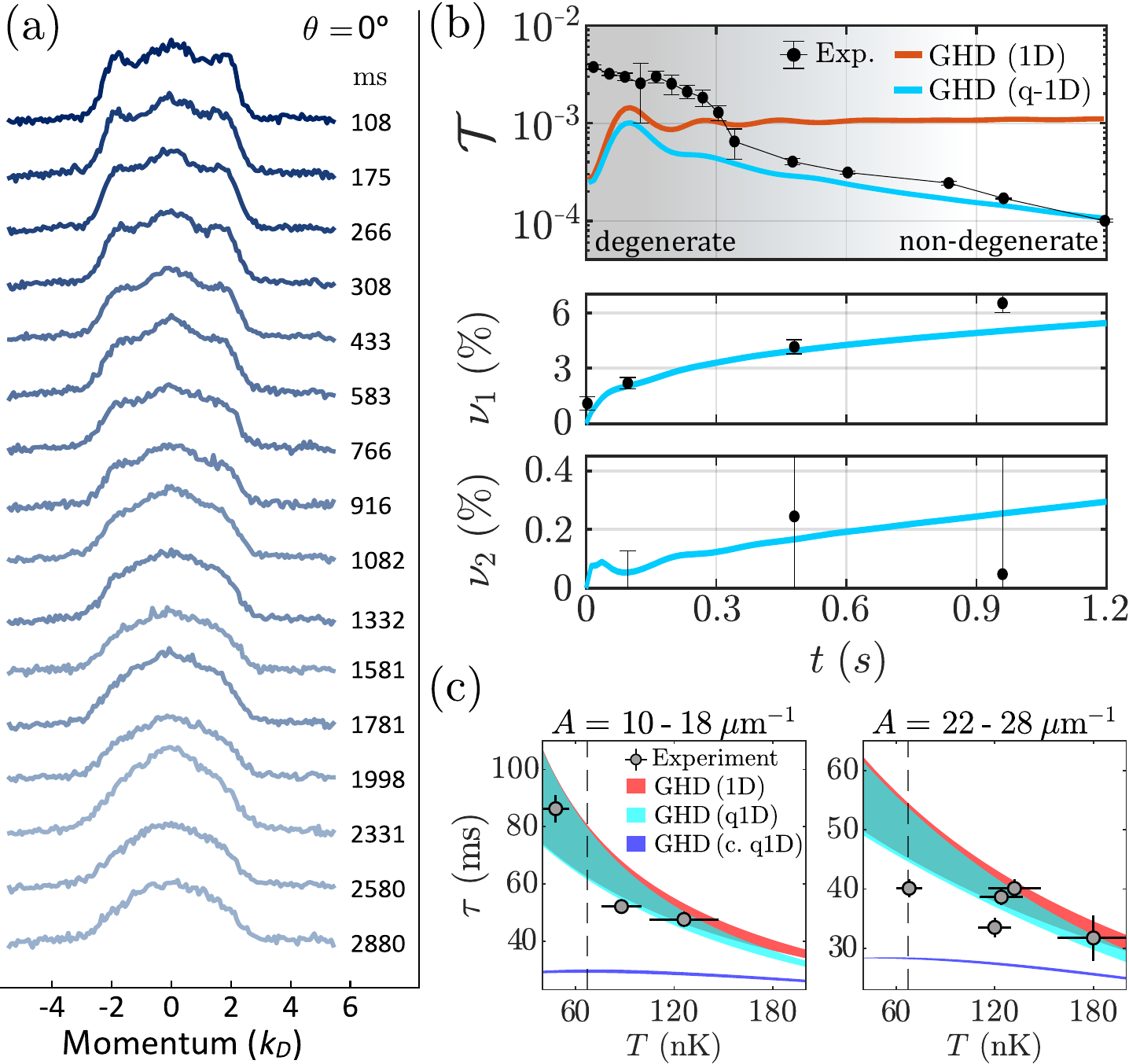}
    \phantomsubfloat{\label{fig:integrability_breaking_a}}
    \phantomsubfloat{\label{fig:integrability_breaking_b}}
    \phantomsubfloat{\label{fig:integrability_breaking_c}}
    \vspace{-2\baselineskip}
    
	\caption{{\bf Integrability breaking and thermalization.} (a) Measured evolution of momentum distribution in a 1D dipolar quantum Newton's cradle. Reproduced from Ref.~\onlinecite{tang2018}. (b) Thermalization of quasi-1D quantum Newton's cradle~\cite{10.21468/SciPostPhys.9.4.058}, quantified as distance of measured momentum distribution to Gaussian fit $\mathcal{T}$, and population of 1st and 2nd excited states of the transverse confining potential $\nu_{1,2}$. Experimental measurements are compared to 1D and quasi-1D GHD, where $\mathcal{T}$ is calculated using rapidity distribution. Reproduced from Ref.~\onlinecite{PhysRevLett.126.090602}. (c) Measured relaxation time $\tau$ of single momentum mode in a 1D box across several temperatures $T$ and mode amplitudes $A$. Agreement with predictions of both 1D and quasi-1D GHD indicate an emergent Pauli blocking of transverse excitations. Reproduced from Ref.~\onlinecite{PhysRevX.12.041032}.}
	\label{fig:integrability_breaking}
\end{figure}

Without controlled theoretical methods, much of what we know about the hydrodynamics of nearly integrable systems comes from experiment. 
A particularly appealing case is that of weakly broken one-dimensionality. 
%
In experiments with ultracold gases, a 1D system is realized by tightly confining the atoms along two transverse directions; if the energy spacing of the transverse potential $\hbar \omega_\perp$ far exceeds any internal energy scales of the gas, then the system can to a good approximation be considered 1D~\cite{PhysRevLett.87.130402, PhysRevLett.87.160405, PhysRevLett.105.265302}.
The dimensional crossover~\cite{guo2023experimental}, or quasi-1D~\cite{Gerbier_2004}, regime can be accessed when the collisional energy per atom exceeds $\hbar \omega_\perp$ facilitating the excitation of atoms in the transverse confining potential.

The emergent hydrodynamics and thermalization in the presence of a weak crossover were studied in the quantum Newton's cradle setup of Refs.~\cite{10.21468/SciPostPhys.9.4.058, PhysRevLett.126.090602}.
By tuning the initial Bragg pulse sequence to impart an average kinetic energy of $\sim 0.9 \hbar \omega_\perp$ onto the atoms, the system was brought to the threshold of dimensional crossover. 
Further fine-tuning of the integrability breaking was achieved by varying the initial temperature and chemical potential; increasing either results in higher energy per atom, whereby thermalization occurs faster.
The ensuing dynamics exhibited a two-stage relaxation: an initial, rapid relaxation dominated by dephasing effects followed by a slower approach towards a thermal distribution.
Notably, even a small fraction of atoms partaking in integrability-breaking scatterings was sufficient to drastically change dynamics.
In order to describe the relaxation dynamics, particularly the latter stage, a multicomponent model~\cite{yang:1967, PhysRevLett.20.98} featuring a Boltzmann-type collision integral in the GHD equation was constructed~\cite{PhysRevLett.126.090602}.
Direct comparisons between observations and GHD were complicated, as the experiment measured the distribution of atomic momenta, rather than that of rapidities.
However, at long evolution times, where the gas was sufficiently close to the non interacting regime for the two distributions to coincide, the collisional model correctly captured the observed dynamics whereas conventional GHD failed.
Furthermore, the fraction of atoms in transverse excited states was measured via band-mapping techniques exhibiting good agreement with the collisional model (Fig.~\ref{fig:integrability_breaking_b}).

\subsubsection{Pauli blocking of diffractive scattering processes}

Integrability-breaking scattering processes, such as the transverse state-changing collisions, create quasiparticles and quasiholes~\cite{2020arXiv200411030D}; 
the scattering rate depends on the available density of states for these.
In the bosonic Lieb-Liniger model, whose quasiparticles are fermionic, the scattering processes can even be Pauli blocked if the rapidities of outgoing scattering states are already occupied.
This scenario was realized in the experiment of Ref.~\cite{PhysRevX.12.041032}:
By confining a quasicondensate in a 1D box on an Atom Chip~\cite{Tajik19}, a Fermi sea at low rapidities could be established across the entire system.
Next, the box bottom was quenched from a cosine-shape to flat, preserving the occupation of low rapidities (unlike the Bragg pulse quench of the quantum Newton’s cradle), and the subsequent evolution of the atomic density was monitored. 
The measurements agreed remarkably well with GHD predictions, even for temperatures up to a few times the transverse level spacing (Fig.~\ref{fig:integrability_breaking_c}), i.e., far from conventional limits of one-dimensionality~\cite{PhysRevA.83.021605}.
Indeed, owing to the emergent Pauli blocking, the Boltzmann collision integral of the quasi-1D model~\cite{PhysRevLett.126.090602} vanishes:
the suppressed scattering thus witnesses the fermionic statistics and also demonstrates that integrability in real systems can be surprisingly robust.

\subsubsection{Dipole-dipole interactions}

Cold-atom experiments often involve atomic species with appreciable dipole-dipole interactions. These, being long-range rather than contact, break integrability. In ``optical-lattice'' experiments that realize arrays of 1D tubes, dipolar interactions give rise to interactions, not just within a tube, but between tubes. These effects have been studied in depth in strongly dipolar dysprosium gases~\cite{PhysRevA.107.L061302}. The accessible dipolar interaction strengths cause resolvable, but quantitatively modest, changes to the equilibrium momentum and rapidity distributions of these gases: to leading order, they renormalize the parameters of the Lieb-Liniger model. However, their effect on dynamics is more pronounced.

An experimental study of dynamics in the presence of DDIs was conducted on the same setup using a quantum Newton’s cradle protocol~\cite{tang2018} (Fig.~\ref{fig:integrability_breaking_a}).
Similar to the experiment of Ref.~\cite{10.21468/SciPostPhys.9.4.058}, the relaxation of dynamics occurred at two different time-scales; one being set by dephasing processes following anharmonicity of the 1D potential and inhomogeneity across the lattice, the other being thermalization due to integrability breaking.
The time-scale of the latter could be tuned by changing the DDI strength, thus enabling a systematic study of perturbations to integrability.
Interestingly, the dynamics of the momentum distribution observed in the experiments of \cite{tang2018} and \cite{10.21468/SciPostPhys.9.4.058} were very similar, despite the mechanism of integrability breaking being very different.
To uncover how different scattering processes affect the rapidity distribution, one would need to measure its evolution.
However, both of these experiments preceded the development of the rapidity distribution measurement~\cite{wilson2019observation}.





\subsection{Correlations, fluctuations and large deviations}
\label{secFCS}

\begin{figure}[t!]
	\centering 
	\includegraphics[width=.99\linewidth]{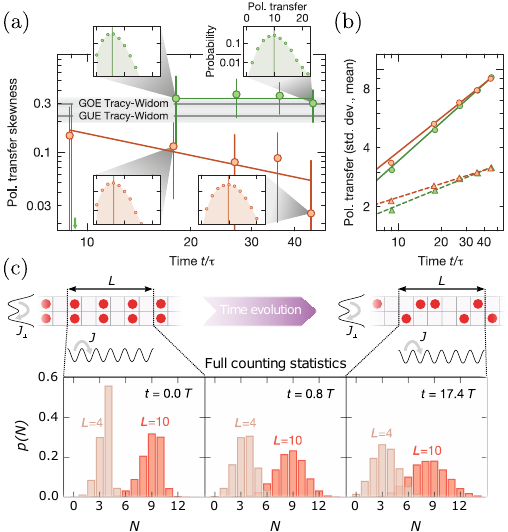}
 
	\caption{{\bf Anomalous transport and full-counting statistics.} (a) Dynamics of the distribution of polarization transfer starting from a  domain-wall initial state for the Heisenberg quantum spin model, observed from quantum gas microscopy. In 2D, the model is non-integrable and  the distribution becomes symmetric at late times (green), whereas the 1D distribution is anomalous and asymmetric with a skewness of 0.33(8). Insets: probability distributions of the polarization transfer. (b)  The mean (circles) of the polarization transfer is consistent with a power-law $t^{1/z}$ (solid lines) with exponent $1/z = 0.67(1)$ in 1D; $1/z = 0.60(2)$ in 2D.  The standard deviation (triangles) features another characteristic transport exponent which agrees with the extracted power-law (dashed lines) exponent, $\beta = 0.31(1)$ in 1D; $\beta = 0.24(1)$ in 2D. In 1D, those experimental results are consistent with the KPZ universality class $z=3/2$ and $\beta=1/3$. Reproduced from Ref.~\onlinecite{wei2022quantum}. (c) Full-counting statistics far-from-equilibrium in ladder systems with tunable rung coupling. Multiple copies of large homogeneous ladder systems are realized using an optical superlattice in the $y$-direction and a simple lattice in the $x$-direction, with a charge-density wave (CDW) initial state.  Adjusting the rung coupling allows to smoothly tune between integrable uncoupled 1D systems of hardcore bosons and a fully-coupled ladder system with non-integrable (chaotic) dynamics. After quenching the system to large tunnel couplings, the CDW rapidly evolves into a state with uniform filling and slowly growing subsystem fluctuations. The full distribution $p(N)$ of the total particle number $N$ in subsystems of length $L$ is measured using single-site resolution. Reproduced from Ref.~\onlinecite{wienand2023emergence}.}
	\label{fig:FCS}
\end{figure}

GHD as originally formulated is a theory of how expectation values evolve after a quantum quench. As we have seen, it can immediately be extended to a theory of two-point correlations by applying hydrodynamic linear response theory. In purely diffusive systems, $j_n=-\mathfrak D_n^i\partial_x q_i + \ldots$, there is a far-reaching framework that extends linear response to all higher-order correlations, response functions and fluctuations at the diffusive scale $t\propto x^2$, fixing their long-time, large-distance behaviors solely from the hydrodynamic data $\mathfrak D_n^i$: macroscopic fluctuation theory (MFT)~\cite{RevModPhys.87.593}. 
Extending MFT to the Euler scale $t \propto x$, by expressing large-scale fluctuations in terms of the Euler-scale data $A_n^i = \partial \j_n[q]/\partial q_i$, remained an open question until very recently. In the past few years, techniques building on GHD have led to rapid progress in this direction~\cite{10.21468/SciPostPhys.5.5.054,fava2021hydrodynamic,alba2019entanglement,fava2023divergent,PhysRevLett.131.140401}, including an extension of MFT to the Euler scale~\cite{DoyonMyers,10.21468/SciPostPhys.8.1.007,PhysRevLett.131.027101,doyon2023ballistic} that subsumes previous model-specific results~\cite{levitov1993charge,doi:10.1063/1.531672, bernarddoyon,1742-5468-2016-6-064005}. 
%
Higher-order correlation functions are increasingly experimentally accessible, and can yield striking signatures of integrable dynamics even in cases where such signatures are absent in linear response. For specificity, we will focus on so-called full counting statistics (FCS)~\cite{levitov1993charge,doi:10.1063/1.531672,PhysRevB.51.4079,PhysRevB.67.085316,PhysRevLett.110.050601, TOUCHETTE20091, Cherng_2007, FCSGritsev, nphys941, 1742-5468-2016-6-064005, garrahan2018aspects, Essler_xxz_fcs, Essler_Ising_fcs, DoyonMyers, 10.21468/SciPostPhys.8.1.007,PhysRevLett.131.140401} and the related large-scale correlation functions, but we note that similar ideas have been used to compute, e.g., the pump-probe response of integrable systems~\cite{fava2021hydrodynamic,fava2023divergent}, where indeed strong signatures of integrability are seen.

\subsubsection{Full counting statistics}

FCS has its origin in mesoscopic physics, as a quantity characterising the non-equilibrium properties of charges hopping through an impurity. It has been used with slightly different but related meanings since. FCS can be defined, for either classical or quantum systems in one dimension, through the following protocol. Consider an infinitely long system partitioned into semi-infinite regions $A$ and $B$ (with interface say at $x=0$), with a conserved charge $\hat Q = \sum_x \hat q(x)$. Define $\hat Q_A = \sum_{x \in A} \hat q(x)$~\footnote{We have left it ambiguous how to treat terms that straddle the boundary between $A$ and $B$, since these will not be important at late times.}.  A system is prepared in some generic state $\rho$, and $\hat Q_A$ is measured at time $t = 0$, yielding the outcome $Q_A(0)$. The system is then evolved to a time $t$ and $\hat Q_A$ is measured again, giving the outcome $Q_A(t)$. These two (correlated) outcomes determine a classical random variable $\Delta Q \equiv Q_A(t) - Q_A(0)$. One is interested in the {\em fluctuations of $\Delta Q$}, and in particular their universal features as $t$ is made large. For this purpose, the object of interest in FCS is the probability distribution $P(\Delta Q)$, or, equivalently, its cumulant generating function, $f(\lambda) \equiv \log \langle e^{i \lambda \Delta Q} \rangle$, where $\langle \cdot \rangle$ denotes an expectation value over the ensemble of initial states. The FCS is the quantity $F(\lambda) = t^{-\alpha} f(\lambda)$, where $\alpha$ is chosen so that the result has a finite limit $t\to\infty$; under generic conditions such as current-current correlations decaying quickly enough  in time, $\alpha=1$. This limit encodes the ``large-deviation theory" of the charge transport fluctuations. Instead of using the two-measurement protocol described here, one can instead directly measure $f(\lambda)$ by using an ancilla to measure the flow of charge across the link~\cite{doi:10.1063/1.531672, samajdar2023quantum}; the protocols are expected to be equivalent, at least in their late-time asymptotics. In classical systems, one can more straightforwardly relate the transferred charge $\Delta Q = \int_0^t dt' j(t')$, where $j$ is the current of the relevant charge across the partition at $x=0$ between the two regions being considered ($j=j_{n_*}$ for one of the available currents $n_*$). There is no difficulty in generalizing to $Q_{A'}(t)-Q_A(0)$ for, say, $A'$ the region $A$ translated by a distance $x\propto t$, and then $\Delta Q = \int_{(0,0)}^{(x,t)} d\vec{s}_\perp\cdot \vec{j}(\vec{s})$. For instance, one may also look at the total charge $\Delta Q$ within a finite region of length $x$ of the system, and its universal features as $x$ is made large.

In the context of GHD, FCS has been measured experimentally (Fig.~\ref{fig:FCS}) starting from far-from-equilibrium initial states using quantum microscopy to study superdiffusive transport in the Heisenberg spin chain~\cite{wei2022quantum}; in coupled ladder systems, where the rung coupling can be used to tune integrability-breaking~\cite{wienand2023emergence}; and on superconducting qubit devices to reveal deviations from KPZ scaling in Heisenberg quantum spin chains~\cite{rosenberg2023dynamics}. 

\subsubsection{Ballistic fluctuation theory}

At the Euler scale, FCS can be computed using the formalism developed in Refs.~\cite{DoyonMyers, 10.21468/SciPostPhys.8.1.007}, which applies to homogeneous states, or that developed in \cite{doyon2023ballistic,PhysRevLett.131.027101} more generally, for inhomogeneous states with non-trivial Euler-scale evolution. We briefly outline the strategy of the former in the case of spatially homogeneous initial states. The Euler-scale current is a classical variable, and one replaces $i\lambda \to \lambda$, so that $f(\lambda) = \log \langle \exp(\lambda \int_0^t dt' j(t')) \rangle$. To compute this, one can re-weight configurations by the exponential of the time-integrated current. This strategy is familiar, e.g., from the theory of macroscopic fluctuations in Markov chains~\cite{lebowitz1999gallavotti}, where an appropriate re-weighting is done at each time-step, and in conformal field theory \cite{1742-5468-2016-6-064005}, where by chirality the time-integrated current is the total charge on a large interval. In general, the evolution of this ``$j$-biased'' ensemble would not be tractable; however, at the Euler scale with $t\to\infty$, and assuming fast enough correlation decay so that $\alpha=1$, a small change of $\lambda$ can be evaluated by a linear response argument, projecting the current onto the conserved quantities, $\delta j_{n_*} = A_{n_*}^i \delta q_i$. This makes the biased ensemble a GGE $e^{-\beta_i(\lambda)Q_i}$, and allows one to use the methods of GHD, or more generally of Euler hydrodynamics, to evaluate the FCS. In the general case of the FCS for the current $(0,0)\to (x,t)$, ensembles biased by strengths $\lambda$ satisfy a {\em flow equation} $d \beta_i/d\lambda = \mathrm{sgn}(x\mathbf{1} - At)_{n_*}^i$,  and one can derive the asymptotic ($x\propto t$ large) relation
$f(\lambda) \sim \int_0^\lambda d\lambda' (t\langle j \rangle_{\lambda'} - x\langle q \rangle_{\lambda'})$, where the expectation values inside the integral are taken in ensembles biased to $\lambda'$ \cite{DoyonMyers,De_Nardis_2022}.
This relation expresses the FCS $F(\lambda)$ purely in terms of hydrodynamic data. This gives not only exact results in integrable models, but is a general procedure applicable once the Euler hydrodynamic data is known; for instance it also reproduces exact results in stochastic exclusion processes (see \cite{MyersThesis}).

In Refs.~\cite{doyon2023ballistic, PhysRevLett.131.027101} these considerations were generalized to spatially fluctuating initial states. An analog to the macroscopic fluctuation theory but for ballistic transport, and in particular for integrable systems, was developed and named the ballistic macroscopic fluctuation theory (BMFT). Strikingly, BMFT allows one to compute consequences of integrable dynamics such as the emergence (and persistence) of long-range correlations after a quantum quench. One finds that, if the model has non-trivial ballistic transport (that is, admits at least two different hydrodynamic velocities), then any non-stationary state will give rise to algebraically small correlations at ballistic-scale space-time, $\langle o_1(\ell x_1,\ell t_1)\cdots o_n(\ell x_n,\ell t_n)\rangle \sim \ell^{-1} S_{o_1,\ldots,o_n}(x_1,\ldots,x_n)$. The BMFT predicts that the function $S_{o_1,\ldots,o_n}(x_1,\ldots,x_n)$ is universal---solely determined by the hydrodynamic data $A$---and non-zero if the initial state is locally relaxed with long wavelengths $\ell$ and the model is truly interacting (that is, the matrix $A$ is not a constant). The ballistic spreading of correlations after a quench was recently experimentally observed~\cite{wienand2023emergence}, although in the free-fermion limit where it can be understood using more elementary methods.

\subsubsection{Phase fluctuations and exponential decay of correlations}

Surprisingly, one can use large-deviation methods to access the dynamical behavior of other observables which are not usually considered in standard hydrodynamics. The idea is to relate these observables to fluctuations of appropriate conserved quantities. One family of observables are phase fields such as the sine-Gordon field, which are experimentally accessible (see Sec.~\ref{intdensityFCS} and Sec.~\ref{intapp}). The sine-Gordon field is related to a ``topological charge", and ballistic fluctuation theory predicts its full large-scale fluctuations in space-time \cite{VecchiosineGordon}. Another family are ``order parameters" which do not overlap with hydrodynamic modes: one may extract their {\em exponential decay} \cite{DoyonMyers,DelVecchioHydroXX,VecchiosineGordon}. This breaks the conventional wisdom which says that hydrodynamic theory is associated with algebraic asymptotic expansions. One predicts that the inverse of the FCS $1/F(\lambda)$ is the correlation time for the two-point correlation function. This is obtained either via Jordan-Wigner transformations such as in the XX chain~\cite{DelVecchioHydroXX}, reproducing results obtained by first-principle analytical means~\cite{PhysRevLett.70.1704,gohmannPrivateComm}; or for vertex operators, which are exponentials of non-compact variables such as, indeed, (de-compactified) phase fields described by the sine-Gordon model~\cite{VecchiosineGordon}, where up to now traditional integrability techniques have failed. Interestingly, in the latter case the predicted exponential decay is in contrast with semi-classical calculations~\cite{sachdev_young,PhysRevLett.95.187201}. The ballistic fluctuation theory appears to account for the exact hydrodynamic modes formed by coherent structures allowed by integrability, and exponential decay is obtained because the topological charge is distributed over a continuum of quasi-particles instead of being carried by an ``isolated" hydrodynamic mode; this is thus a signature of integrability. In non-integrable models, where similar observables exist, the decay is expected to be weaker \cite{PhysRevB.106.205151}.

Currently the technology for fluctuations at the Euler scale, including long-range correlations, inhomogeneous setups, phase fluctuations, and certain exponential decay, is rather advanced, but independent numerical, experimental or analytical confirmations~\cite{gohmann2023thermal} are challenging. 
Further, hydrodynamic expressions for many observables, such as the fundamental field of the Lieb-Liniger model which contains information about the physical momentum distribution, are still to be worked out. 
Relating these fluctuations to quantum entanglement entropies out of equilibrium using these ideas
\cite{alba2017entanglement,alba2019entanglement,PhysRevLett.131.140401}, and understanding the entanglement / fluctuation correspondence in interacting models~\cite{PhysRevLett.102.100502,PhysRevB.85.035409,Calabrese_2012,del2023entanglement}, remain open questions.

\subsubsection{Anomalous FCS beyond Euler scale}

While the ballistic fluctuation theory framework applies to Euler-scale dynamics, some of the most surprising aspects of FCS in integrable systems lie beyond the Euler scale. A general framework for these is so far missing. 
For instance, one may ask about quantum effects \cite{Bernard_2021}, 
diffusive and higher-order effects, etc. 
These new regimes have already been partially explored by a combination of experiments, numerics, and exact solutions of integrable cellular automata. 
These explorations have led to the unexpected discovery of a new, strongly non-gaussian regime of FCS in a family of integrable models, including the easy-axis XXZ spin chain. We briefly describe these developments, focusing on the simple case of the large-$\Delta$ limit of the XXZ spin chain~\cite{PhysRevLett.128.090604, 2022arXiv220309526G}. 

For simplicity let us consider an initial state that is a high-contrast magnetic domain wall, i.e., $\rho = \bigotimes_i e^{-\mu \,\mathrm{sgn}(i) S^z_i}$, with $\mu \gg 1$. In the limit $\mu \to \infty$, all spins to the left (right) of the origin, point along $\downarrow (\uparrow)$. In terms of quasiparticles, this state can be seen as a single giant quasiparticle occupying one half of the system; for $\Delta > 1$ this quasiparticle is immobile in the thermodynamic limit. At large but finite $\mu$, each side of the domain wall is sprinkled with a low density of minority spins, which are magnons (Sec.~\ref{secXXZ}). The domain wall moves diffusively through collisions with magnons: when a magnon moves right across the domain wall, the domain wall shifts two steps left, and vice versa. In this dilute limit, each crossing can be treated separately. At each crossing event, the magnetization transfer across the origin changes as follows: \emph{when the domain wall moves away from the origin, magnetization is transferred to the left; when it moves toward the origin, magnetization is transferred to the right}. Thus the history of magnetization transfer in this limit follows a Dyck path, i.e., a random walk that reflects when it hits the origin. The histogram of magnetization transfer is, accordingly, that of a reflecting random walk: $P(\Delta Q) \sim 2/\sqrt{\pi D t} \exp(-\Delta Q^2/(D t)) \Theta(\Delta Q)$, where $\Theta(t)$ is the Heaviside step function. 

This distribution is quite different from the FCS of standard diffusive systems~\cite{Derrida,RevModPhys.87.593,mcculloch2023counting}. First, its mean and standard deviation scale the same way with time, as $\sqrt{t}$, and are comparable at all times. Second, its skewness, kurtosis, and other normalized cumulants approach large nonzero values at late times, again in contrast to standard expectations. These phenomena are essentially due to the fact that charge transport is governed by the motion of a \emph{single} random walker, rather than an ensemble of independent walkers. It is a familiar observation that shot noise is enhanced when the charge carriers are larger (so charge transport is more ``granular''): the parametric enhancement of magnetization noise in the present context is a diagnostic of the mechanism whereby spin transport is due to giant quasiparticles.

This result is asymptotically true for \emph{any} $\mu \neq 0$: on large enough scales any domain wall can be characterized by its average contrast, which only renormalizes the diffusion constant above. The general formula is~\cite{2022arXiv220309526G}
\begin{equation}
P(\Delta Q) = \int_0^\infty dx \frac{\exp \left(-\frac{x^2}{4 D t}-\frac{x \left(\frac{\Delta Q}{x}-m\right)^2}{2 \left(1-m^2\right)}\right)}{\sqrt{2 \pi^2 (1-m^2)} \sqrt{D t x}}, 
\end{equation}
where $m= \tanh \mu$. This distribution is non-Gaussian even in the equilibrium case $m=0$, where it characterizes magnetization transfer at the scale $\Delta Q \sim t^{1/4}$. 
This equilibrium limit was first observed numerically in Ref.~\cite{PhysRevLett.128.090604}, and independently derived, using different methods, for single-file diffusion in Ref.~\cite{krajnik2022exact}. Remarkably, this result is nonanalytic at $\Delta Q = 0$. When the system is biased slightly out of equilibrium, with a domain wall of contrast $\epsilon$, the growth of variance crosses over from equilibrium to the nonequilibrium behavior on a timescale $t(\epsilon) \sim 1/\epsilon^4$.


\subsection{Rigorous proofs of GHD}

Recent results have also established the validity of the GHD equations at various levels of rigor. In particular, the ``rate-equation''~\eqref{eqVeff} was proved (under various mild assumptions) using combinatorial techniques~\cite{10.21468/SciPostPhys.6.2.023}, the existence of a self-conserved current~\cite{10.21468/SciPostPhys.9.3.040}, form factors~\cite{CortesCubero2019}, and algebraic Bethe ansatz techniques~\cite{Balasz,PhysRevLett.125.070602,10.21468/SciPostPhys.8.2.016}. Up to the assumption of local relaxation, taken together those results firmly establish the validity of GHD at the Euler scale. 
We refer the reader to the recent reviews~\cite{Cubero_2021,Borsi_2021} for more details. In certain classical systems strong mathematical results are available concerning the emergence of the GHD equation \cite{Boldrighini1983,EL2003374,ferrari2022hard}, and recent works have also explored ab-initio derivations in quantum systems, forgoing the assumption of local relaxation, starting from kinetic theory and semi-classical principles~\cite{PhysRevLett.128.190401,doyon2023ab}. But much work remains to be done to establish the emergence of GHD starting from generic non-equilibrium states in the quantum case. Proving the emergence of hydrodynamic equations from microscopic, deterministic dynamics is in general an extremely challenging problem, however there is the hope that with integrability techniques, this will be possible for GHD. Of much interest will be to prove the form of higher-order terms including diffusion constants and quantum corrections.

\section{Conclusions}

Many experimentally relevant one-dimensional systems are close to integrable limits, and cannot be described by conventional hydrodynamics. Instead, their dynamics can be captured by GHD, which provides a universal and quantitative framework to study far-from-equilibrium transport in 
systems with infinitely-many conserved quantities and long-lived quasiparticle excitations. 
We have reviewed the recent successes of GHD, from computing various transport coefficients such as diffusion constants that were previously thought to be completely out of reach, to quantitative descriptions of the far-from-equilibrium dynamics of trapped Bose gases in arbitrary smooth time- and space-dependent potentials. These theoretical results sparked a new generation of experiments testing the predictions of GHD, which in turn also led to new challenges for theory, including the discovery of new regimes of anomalous transport, the emergence of hydrodynamics in systems with a relatively small number of particles where hydrodynamics might naively not apply, and the measurement full counting statistics. 

Only a few years after its discovery, GHD is now part of the standard toolkit of non-equilibrium dynamics in one dimension. As we have outlined in Sec.~\ref{secOpenQuestions}, many open questions remain---including establishing the emergence of GHD microscopically more rigorously, understanding anomalous regimes of transport, further extending beyond their traditional territory the predictions from hydrodynamic principles such as the exponential decay of correlation functions and the large-scale behavior of quantum entanglement, systematically incorporating the effects of integrability-breaking perturbations needed to describe realistic systems and the quantum effects emerging at small temperatures, and fully understanding the GHD phenomenology including the possible emergence of turbulence-like behaviors. Another important frontier which is up to now little explored is that of higher-dimensional integrable systems, where one may hope to provide a theory for certain turbulent surface water waves. We expect that future works will shed light on these important questions, and will further strengthen the importance of GHD, and the reach of hydrodynamic theory more generally, to describe non-equilibrium many-body classical and quantum dynamics in one and perhaps higher dimensions.

\begin{acknowledgments}

We thank Monika Aidelsburger, Utkarsh Agrawal, Vincenzo Alba, Alvise Bastianello, Denis Bernard, M. Joe Bhaseen, Immanuel Bloch, Sounak Biswas, Thibault Bonnemain, Vir Bulchandani, Olalla Castro Alvaredo, Jean-S\'ebastien Caux, Giuseppe Del Vecchio Del Vecchio, J\'er\^ome Dubail, Joseph Durnin, Simon Karch, Jacopo De Nardis, Michele Fava, Aaron Friedman, Paolo Glorioso, Rosemary J. Harris, Friedrich H\"ubner, David Huse, Enej Ilievski,  Alexander Impertro,  Vedika Khemani, M\'arton Kormos, Javier Lopez-Piqueres, Igor Mazets, Catherine McCarthy, Ewan McCulloch, Marko Medenjak (deceased), Joel Moore, Jason Myers, Vadim Oganesyan, Sid Parameswaran, Gabriele Perfetto, Tomaz Prosen, Marcos Rigol, Subir Sachdev, Tomohiro Sasamoto, Christian Schweizer, Hans Singh, Herbert Spohn, Brayden Ware, Julian Wienand, Takato Yoshimura and Marko Znidaric for collaborations and/or numerous discussions on topics related to GHD. R.V. acknowledges support from the US Department of Energy, Office of Science, Basic Energy Sciences, under award No. DE-SC0023999, and the Alfred P. Sloan Foundation through a Sloan Research Fellowship.  J.S. ans F.M. acknowledge support from the European Reserach Council: ERC-AdG: Emergence in Quantum Physics (EmQ). B.D. was supported by the Engineering and Physical Sciences Research Council (EPSRC) under grant EP/W010194/1.

\end{acknowledgments}

\bibliography{ghdbib}

\end{document}